\documentclass[10pt,prl,twocolumn,showpacs,floatfix,superscriptaddress]{revtex4-1}
\usepackage{amsmath,amsbsy,graphics,epsfig,mathrsfs}
\usepackage{bm}
\usepackage[]{color}
\usepackage{graphicx,epstopdf}
\usepackage{hyperref}
\usepackage{cancel}

\begin{document}

\title{Realization of anisotropic compass model on the diamond lattice of Cu$^{2+}$ in CuAl$_2$O$_4$}

\author{S.~A.~Nikolaev}
\affiliation{National Institute for Materials Science, MANA, 1-1 Namiki, Tsukuba, Ibaraki 305-0044, Japan}
\email{saishi@inbox.ru}

\author{I.~V.~Solovyev}
\affiliation{National Institute for Materials Science, MANA, 1-1 Namiki, Tsukuba, Ibaraki 305-0044, Japan}
\affiliation{Department of Theoretical Physics and Applied Mathematics, Ural Federal University, Mira St. 19, 620002 Yekaterinburg, Russia}
\email{SOLOVYEV.Igor@nims.go.jp}

\author{A.~N.~Ignatenko}
\affiliation{Institute of Metal Physics, S.Kovalevskoy St. 18, 620108 Yekaterinburg, Russia}

\author{V.~Yu.~Irkhin}
\affiliation{Institute of Metal Physics, S.Kovalevskoy St. 18, 620108 Yekaterinburg, Russia}

\author{S.~V.~Streltsov}
\affiliation{Institute of Metal Physics, S.Kovalevskoy St. 18, 620108 Yekaterinburg, Russia}
\affiliation{Department of Theoretical Physics and Applied Mathematics, Ural Federal University, Mira St. 19, 620002 Yekaterinburg, Russia}
\email{streltsov@imp.uran.ru}

\date{\today}

\begin{abstract}
Spin-orbit (SO) Mott insulators are regarded as a new paradigm of magnetic materials, whose properties are largely influenced by SO coupling and featured by highly anisotropic bond-dependent exchange interactions between the spin-orbital entangled Kramers doublets, as typically manifested in $5d$ iridates. Here, we propose that a very similar situation can be realized in cuprates when the Cu$^{2+}$ ions reside in a tetrahedral environment, as in spinel compounds. Using first-principles electronic structure calculations, we construct a realistic model for the diamond lattice of the Cu$^{2+}$ ions in CuAl$_2$O$_4$ and show that the magnetic properties of this compound are largely controlled by anisotropic compass-type exchange interactions that dramatically modify the magnetic ground state by lifting the spiral spin-liquid degeneracy and stabilizing a commensurate single-$\boldsymbol{q}$ spiral.
\end{abstract}

\maketitle

\par \emph{Introduction}. The relativistic spin-orbit (SO) interaction manifests itself in numerous spectacular phenomena that constitute a large area of modern condensed matter physics. Besides the explosive growth of investigations on various topological aspects of matter~\cite{topology}, there is a strong interest in unconventional types of SO assisted magnetism, which can be realized in otherwise conventional transition-metal oxides. The main activity here is focused on iridates, which become the key testbed materials for exploring new SO coupling driven effects~\cite{Kim2008,StreltsovKhomskii}. Apart from a strong SO interaction, splitting the lowest $t_{2g}$ manifold of octahedrally coordinated Ir$^{4+}$ ions into relativistic $j= 3/2$ and $1/2$ states, another important aspect of iridates that makes them unique among $5d$ oxides is a single hole occupancy of this manifold. Since no other interactions are involved, the character of the $t_{2g}$ hole is solely controlled by SO coupling: the hole is accommodated in the subshell of Kramers degenerate $j = 1/2$ pseudospin states, where it experiences the on-site Coulomb interaction $U$, driving the Mott transition. Corresponding exchange interactions are subjected to Anderson's superexchange theory~\cite{Anderson1959}, but since the $j=1/2$ state mixes spin and orbital variables, such a theory is inevitably flavored by strong and bond-dependent (compass) anisotropy~\cite{Khaliullin2009}. This is the key property of iridates, which is largely implicated in various phenomena, such as the antisymmetric Dzyaloshinkii-Moriya (DM) interaction related magnetic phase transition in Sr$_2$IrO$_4$~\cite{BJKimScience2009}, a possible realization of Kitaev spin-liquid states in Na$_2$IrO$_3$~\cite{Chaloupka}, and unconventional spiral order in Li$_2$IrO$_3$~\cite{Biffin2014}.

\par Is such rich unusual physics solely inherent to iridates? In fact, the requirement of ``strong SO coupling'' itself does not seem to be crucial in comparison with two other conditions that could be used in the search for \emph{non}-$5d$ analogs of iridates, namely: (i) a single hole occupancy of the SO active $t_{2g}$ shell, and (ii) large $U$, which turns the system into the Mott regime and thus amplifies the effect of SO coupling. A possible $t_{2g}$ level splitting by crystal field may change the situation quantitatively, but not the concept of the pseudospin itself: the hole will always reside in a Kramers doublet, serving as the basis for pseudospin states~\cite{Khaliullin2009}. Such a situation is indeed realized in $\alpha$-RuCl$_3$~\cite{RuCl3}, which is a $4d$ electron analog of Na$_2$IrO$_3$. Another interesting possibility is to have Cu$^{2+}$ or any other $d^{9}$ ions in the \textit{tetrahedral} environment, which is characterized by a negative $10Dq$ splitting, so that the single hole resides in the high-lying $t_{2g}$ manifold.

\par The well-known example of tetragonally coordinated transition-metal atoms in oxides is the $A$-site spinels $AB_{2}$O$_{4}$. The unusual magnetic properties of spinels have attracted considerable attention in the context of magnetic frustration driven by competing exchange interactions~\cite{Loidl2005,Loidl2006,Loidl2008}. Particularly, a novel spiral spin-liquid state has been predicted on the diamond lattice of transition-metal ions in $AB_{2}$O$_4$, where mean-field calculations for classical spins showed that, if the ratio of exchange parameters between nearest and next-nearest neighbors (hereafter $nn$ and $nnn$, respectively) is greater than $1/8$, a degenerate state of coplanar spin spirals is stabilized with a continuous complex surface of propagation vectors in the reciprocal space~\cite{Trebst2007}. This state is persistent up to sufficiently low temperatures, when degeneracy is ultimately lifted by thermal fluctuations due to an order-by-disorder mechanism~\cite{Villain1980}. Indeed, the unconventional spin order has been reported in CoAl$_2$O$_4$~\cite{Iakovleva2015,Zaharko2011} and MnSc$_{2}$S$_{4}$~\cite{ruegg2017}, where a multi-step ordering process of the spiral spin-liquid was observed. However, low-temperature ordering in highly frustrated magnets is extremely sensitive to various perturbations, and in real systems magnetic frustration can often be relieved when spins are coupled to the lattice via SO interactions. Interestingly, recent theoretical studies have also shown that nontrivial topological spin excitations and charge responses may be realized in antiferromagnets on the diamond lattice with SO coupling~\cite{sekine}. 

\par Among many Cu spinels, the most interesting candidate for studying anisotropic compass interactions is CuAl$_2$O$_4$. (i) It remains cubic in a whole temperature range~\cite{Nirmala2017}. Although SO coupling in CuAl$_2$O$_4$ is about five times weaker than in iridates, it is solely responsible for the atomic $t_{2g}$ level splitting. (ii) While the Curie-Weiss temperature of CuAl$_2$O$_4$ is $-137$~K, the magnetic long-range order does not emerge down to $0.4$~K, suggesting high magnetic frustration. Thus, the issue of exchange anisotropy and its role in relieving magnetic frustration seems to be very important in the physics of CuAl$_2$O$_4$.

\par In this work we propose CuAl$_2$O$_4$ to be a SO Mott insulator with a single unoccupied $j=1/2$ state. Using the state-of-the-art first-principles Wannier functions technique, we derive the corresponding pseudospin model and find a strong competition between $nn$ and $nnn$ isotropic exchange interactions, common to the entire family of $A$-site spinels, which tends to suppress a long-range magnetic order~\cite{Trebst2007}. However, we also reveal exceptionally strong DM interactions that lift the degeneracy of the magnetic ground state stabilizing a single-${\boldsymbol q}$ spiral state. The appearance of anisotropic exchange is well expected in the spinel $Fd\bar{3}m$ space group, but to the best of our knowledge this possibility has never been considered before, neither has the effect of DM interactions been addressed in regard to the magnetic properties in spinel compounds.

\begin{figure}[t!]
\begin{center}
\includegraphics[width=1\columnwidth]{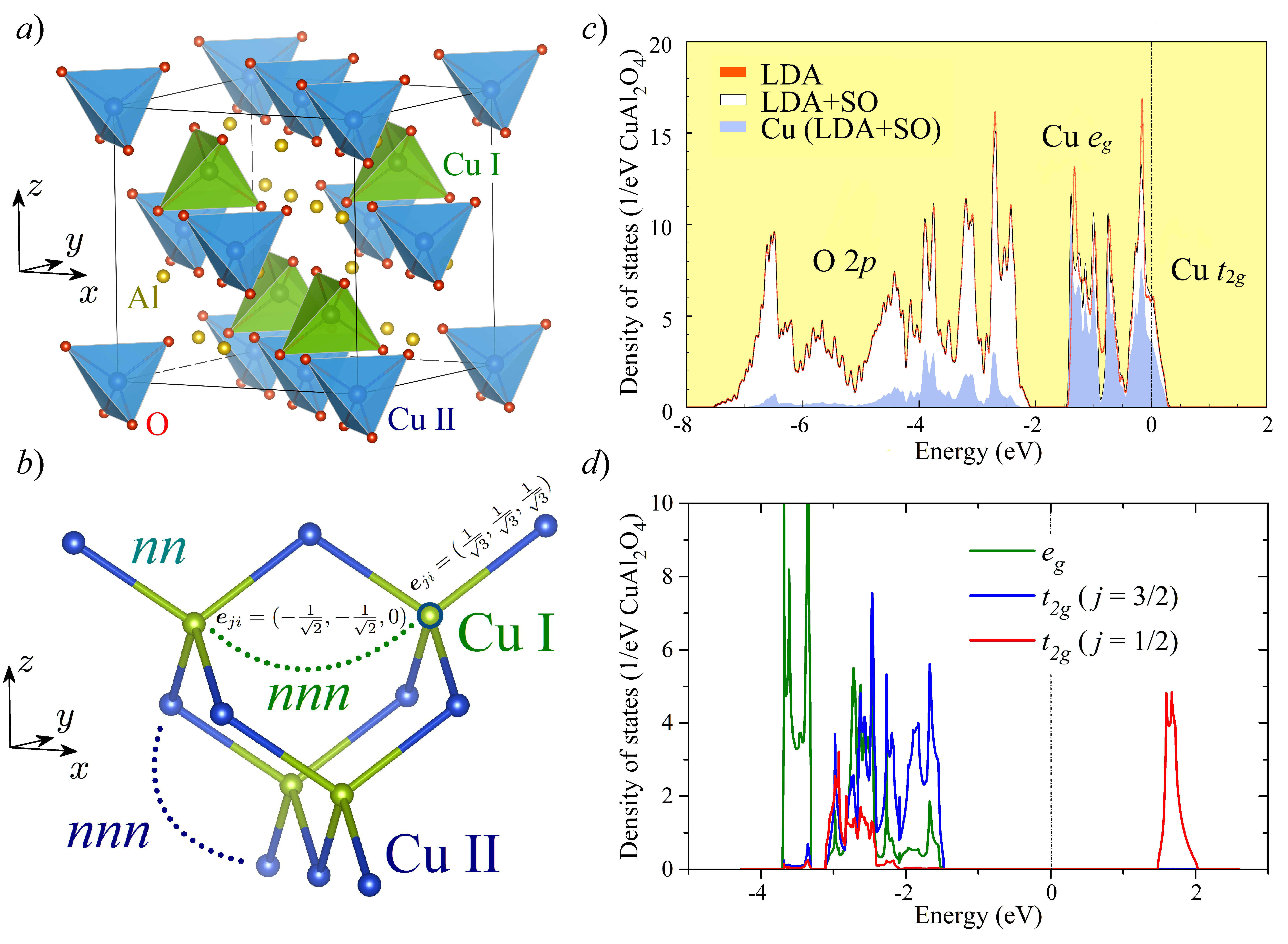}
\end{center}
\caption{$a)$~Crystal structure of CuAl$_2$O$_4$ (generated by Vesta \cite{VESTA}). $b)$~Nearest and next-nearest neighbors on the diamond lattice with two fcc sublattices Cu~I and Cu~II. $c$)~Density of states of CuAl$_2$O$_4$ in LDA with and without SO coupling. $d$)~Partial densities of $e_{g}$ and $t_{2g} (j=1/2$ and $3/2)$ states in the Hartree-Fock approximation for the ferromagnetic solution of the model~(\ref{eqn.ManyBodyH}). The Fermi level is at zero energy.}
\label{fig_cryst}
\end{figure}

\par \emph{Electronic model}. We adopt the experimental structure at $T=40$~K~\cite{Nirmala2017}. The primitive cell contains two Cu sites, which form a diamond lattice with two interpenetrating face-centered cubic (fcc) sublattices I and II shifted relative to each other by one quarter along the cube diagonal~(Fig.~\ref{fig_cryst}). Importantly, the two sublattices can be transformed into one another under inversion symmetry, which in turn is absent in each sublattice, thus allowing for antisymmetric effects.

\par First-principles calculations were carried out by using the norm-conserving pseudopotentials within local density approximation (LDA) as implemented in the Quantum-ESPRESSO package~\cite{qe}. Unlike many other cuprates, falling into the charge-transfer regime~\cite{ZSA}, the electronic structure of CuAl$_2$O$_4$ reveals well isolated Cu $3d$ bands (Fig.~\ref{fig_cryst}$c$) \cite{supp}. The reason is that a strong O-Al hybridization pushes the bonding O $2p$ states away from the Fermi level and reduces the Cu $3d$ bandwidth. In addition, the Cu-O hybridization itself is weaker in the tetrahedral environment (compared to the octahedral one).

\par As a starting point, we construct the effective model in the Wannier basis for the Cu $3d$ bands~\cite{wan90}:
\begin{eqnarray}
\hat{\cal{H}}&=&  \sum_{ij} \sum_{\sigma \sigma'} \sum_{ab}
\left( t_{ij}^{ab}\delta_{\sigma \sigma'} + \Delta^{ab\sigma\sigma'}_{i}\delta_{ij}\right)
\hat{c}^\dagger_{ia\sigma}
\hat{c}^{\phantom{\dagger}}_{jb\sigma'}\nonumber \\
&+&  \frac{1}{2}
\sum_{i}  \sum_{\sigma \sigma'} \sum_{abcd} U_{i}^{abcd}
\hat{c}^\dagger_{i a \sigma} \hat{c}^\dagger_{i c \sigma'}
\hat{c}^{\phantom{\dagger}}_{i b \sigma}
\hat{c}^{\phantom{\dagger}}_{i d \sigma'},
\label{eqn.ManyBodyH}
\end{eqnarray}
\noindent where $\hat{c}^{\dagger}_{ia\sigma}$ is the creation operator of an electron with spin $\sigma$ at site $i$ and orbital $a$. We consider two types of such models constructed for the minimal set of $t_{2g}$ bands and in the full $3d$ basis. $\hat{t}$ in Eq.~(\ref{eqn.ManyBodyH}) are the hopping parameters, $\hat{\Delta}$ is the matrix of the SO interaction (which also includes $10Dq$ splitting of $-$$0.82$ eV for the full $3d$ model), and $U_{i}^{abcd}$ are the screened Coulomb interactions calculated within the constrained random phase approximation (cRPA)~\cite{cRPA,Solovyev2008}. All model parameters are given in the Supplemental Material~\cite{supp}. The $3d$ model is more advanced as it explicitly treats the Cu $e_{g}$ bands. Nevertheless, as we will see below, both models provide similar sets of exchange parameters, when treated within superexchange theory.
\par In the $t_{2g}$ model, $\hat{t}$ can be written in a compact form in terms of the unit vectors $\boldsymbol{e}_{ji} \equiv (e_{ji}^x,e_{ji}^y,e_{ji}^z)$ along the bonds between two Cu sites. In each sublattice there are four $nn$ centrosymmetric bonds that connect Cu~I and Cu II sites and obey the trigonal symmetry. For the sublattice I, the unit vectors are $\boldsymbol{e}_{ji} = (\frac{1}{\sqrt 3},\frac{1}{\sqrt 3},\frac{1}{\sqrt 3})$, $(-\frac{1}{\sqrt 3},-\frac{1}{\sqrt 3},\frac{1}{\sqrt 3})$, $(-\frac{1}{\sqrt 3},\frac{1}{\sqrt 3},-\frac{1}{\sqrt 3})$, and $(\frac{1}{\sqrt 3},-\frac{1}{\sqrt 3},-\frac{1}{\sqrt 3})$ (Fig.~\ref{fig_cryst}$c$), and we obtain $t_{ij}^{ab}=t_{nn}\delta^{ab}$$+$$\Delta t_{nn}(e^{a}_{ji}e^{b}_{ji}$$-$$\frac{1}{3}\delta^{ab})$, where $t_{nn}$$=-$$43$ meV and $\Delta t_{nn}$$= -$$22$ meV. For the sublattice II, we have $\hat{t}^{\mathrm{II},\mathrm{I}}_{ji}=\hat{t}^{\mathrm{I},\mathrm{II}}_{ij}$.


\par  The 12 $nnn$ sites belong to the same sublattice and are separated by the fcc lattice translations $\boldsymbol{e}_{ji}=(\pm\frac{1}{\sqrt{2}},\pm\frac{1}{\sqrt{2}},0)$, $(\pm\frac{1}{\sqrt{2}},0,\pm\frac{1}{\sqrt{2}})$, and $(0,\pm\frac{1}{\sqrt{2}},\pm\frac{1}{\sqrt{2}})$ (Fig.~\ref{fig_cryst}$c$). As they are \textit{not} connected by inversion symmetry, the corresponding transfer integrals contain both symmetric ($S$) and antisymmetric ($A$) parts, $\hat{t}=\hat{t}^{(S)}+\hat{t}^{(A)}$. The $S$-part has the following form: $t_{ij}^{(S)ab}=t_{nnn}\delta^{ab}+\Delta t_{nnn}(e^{a}_{ji} e^{a}_{ji} - \frac{1}{3})\delta^{ab} + \Delta t'_{nnn} (e^{a}_{ji} e^{b}_{ji} -\frac{1}{2} \delta^{ab})$, where $t_{nnn} = 26$ meV, $\Delta t_{nnn} = 141$ meV, and $\Delta t'_{nnn} = 113$ meV. The $A$-part can be expressed as $t_{ij}^{(A)ab}=\varepsilon^{abc}\nu^{c}_{ij}$, in terms of the Levi-Civita symbol $\varepsilon^{abc}$ and the axial vector $\boldsymbol{\nu}_{ij}=\nu[\boldsymbol{n} \times \boldsymbol{e}_{ji}]$, which is perpendicular to the bond $\boldsymbol{e}_{ji}$ and the corresponding face of the cube, where $\nu=28$ meV, $\boldsymbol{n}^{\mathrm{I}}= -\boldsymbol{n}^{\mathrm{II}} = (\mp1,0,0)$ for either $\boldsymbol{e}_{ji} = (0,\frac{1}{\sqrt{2}},\pm \frac{1}{\sqrt{2}})$ or $\boldsymbol{e}_{ji}=(0,-\frac{1}{\sqrt{2}},\mp\frac{1}{\sqrt{2}})$ (for the rest of the bonds one can apply the cyclic permutation of indices 1, 2, and 3). In short, $\hat{t}^{\mathrm{I},\mathrm{I}}_{ij}=[\hat{t}^{\mathrm{II},\mathrm{II}}_{ij}]^{T}$ holds true. While the existence of $\hat{t}^{(A)}$ itself is related to the lack of inversion symmetry in each sublattice, its form is determined by the symmetry of $nnn$ bonds, which can be transformed into themselves by rotating $180^{\circ}$ around $\boldsymbol{n}$.

\par In the $t_{2g}$ model, the matrix elements of $\hat{U}$ obey the Kanamori rules~\cite{Kanamori}: $U=U'+2J_{\rm H}$, where $U$ and $U'$ are, respectively, the intraorbital and interorbital Coulomb interactions, and $J_{\rm H} = 0.7$ eV is the Hund's rule exchange coupling. Since the Cu $3d$ bands are well separated and nearly filled, the cRPA screening is not particularly strong. As a result, the parameter $U$ is large ($5.0$ and $5.4$ eV in the $t_{2g}$ and full $3d$ model, respectively), thus fully justifying the use of superexchange theory~\cite{Anderson1959}.

\par The calculated SO coupling constant, $\xi\sim122$ meV, is comparable to hopping parameters in individual bonds. However, since the number of bonds is large (4 and 12 for $nn$ and $nnn$, respectively), in the ordinary LDA the relativistic $j=3/2$ and $1/2$ states are strongly entangled by $\hat{t}$, leading to nearly identical band structure with and without SO coupling (Fig.~\ref{fig_cryst}$c$). Nevertheless, large $U$ will effectively suppress electron hoppings, thus unveiling the atomic energy scale given by SO coupling. This is clearly seen from the self-consistent Hartree-Fock calculations for the $3d$ model (\ref{eqn.ManyBodyH}): densities of states projected onto the $e_{g}$ and $j=3/2$ and $1/2$ $t_{2g}$ orbitals (obtained by diagonalizing $\hat{\Delta}$) reveal a pure $j=1/2$ character of the unoccupied (hole) states (Fig.~\ref{fig_cryst}$d$). The obtained spin and orbital magnetic moments are about $1/3$ $\mu_{\rm B}$ and $2/3$ $\mu_{\rm B}$, respectively, as expected for the $j=1/2$ states in the undistorted cubic environment~\cite{BJKimScience2009}. Thus, it is totally legitimate to apply the theory of superexchange and derive the exchange interactions for CuAl$_{2}$O$_{4}$ by considering virtual excitations to the unoccupied $j=1/2$ states formed by the highest Kramers doublet of $\hat{\Delta}$, as was previously done for iridates~\cite{Khaliullin2009,SolovyevPRB2015}.

\par \emph{Magnetic interactions}. The corresponding magnetic model is formulated as:
\begin{equation}
 \label{eqn:HS}
\mathcal{H}_{\mathcal{S}} = \sum_{i > j} \boldsymbol{\mathcal S}_{i}
\tensor{J}_{ij} \boldsymbol{\mathcal S}_{j},
\end{equation}
\noindent where $\tensor{J}_{ij}$ is the $3$$\times$$3$ tensor of exchange interactions between $j=1/2$ pseudospins $\boldsymbol{\mathcal{S}}_{i}$ comprising both spin and orbital parts.
\par The resulting exchange interactions obey the symmetry properties of $\hat{t}$ in the $t_{2g}$ model, that is $\tensor{J}_{ij}^{\mathrm{I},\mathrm{II}}=\tensor{J}_{ji}^{\mathrm{II},\mathrm{I}}$ and $\tensor{J}_{ij}^{\mathrm{II},\mathrm{II}}=[\tensor{J}_{ij}^{\mathrm{I},\mathrm{I}}]^{T}$. So, $\tensor{J}_{ij}$ between $nn$ sites is totally symmetric and can be represented in the form $J_{ij}^{\alpha \beta}=J_{nn} \delta^{\alpha \beta} + \Delta J_{nn} (e^{\alpha}_{ji} e^{\beta}_{ji} -\frac{1}{3}\delta^{\alpha \beta})$, where the values of the isotropic $J_{nn}$ and anisotropic $\Delta J_{nn}$ components in the full $3d$ ($t_{2g}$) model are $1.72$ ($2.10$) meV and $0.24$ ($0.21$) meV, respectively. Furthermore, it follows that $\Delta J_{nn}$ is controlled by $J_{\rm H}$ and vanishes when $J_{\rm H}=0$, in agreement with Jackeli-Khaliullin theory~\cite{Khaliullin2009}.

\par The $nnn$ exchange interactions  contain both symmetric and antisymmetric parts, $\tensor{J}_{ij}\equiv\tensor{J}_{ij}^{(S)}+\tensor{J}_{ij}^{(A)}$. The $S$-part can be written as
$J_{ij}^{(S)\alpha\beta}=J_{nnn}\delta^{\alpha\beta}+\Delta J_{nnn}(e^{\alpha}_{ji}e^{\alpha}_{ji}-\frac{1}{3})\delta^{\alpha\beta}+\Delta J'_{nnn}(e^{\alpha}_{ji}e^{\beta}_{ji}-\frac{1}{2}\delta^{\alpha\beta})$, where for the full $3d$ ($t_{2g}$) model $J_{nnn}=0.48$ ($0.45$) meV, $\Delta J_{nnn}=1.20$ ($1.31$) meV and $\Delta J'_{nnn}=0.08$ ($0$) meV. The $A$-part is expressed by DM vectors $\boldsymbol{d}_{ij}=(d_{ij}^x,d_{ij}^y,d_{ij}^z)$ as $\boldsymbol{\mathcal{S}}_i\tensor{J}^{(A)}_{ij}\boldsymbol{\mathcal{S}}_j =
\boldsymbol{d}_{ij}\cdot[\boldsymbol{\mathcal{S}}_{i}\times\boldsymbol{\mathcal{S}}_{j}]$. The vectors $\boldsymbol{d}_{ij}$ are related to the asymmetry of $\hat{t}_{ij}$ and obey the same symmetry properties as $\boldsymbol{\nu}_{ij}$: $\boldsymbol{d}_{ij}^{\mathrm{I}(\mathrm{II})}=d[\boldsymbol{n}^{\mathrm{I}(\mathrm{II})}\times\boldsymbol{e}_{ji}]$, where $d=1.10$ ($1.04$) meV for the full $3d$ ($t_{2g}$) model. Thus, the $nnn$ anisotropic exchange is considerably larger than the isotropic one, and the asymmetry of transfer integrals due to the lack of inversion centers connecting \emph{nnn} sites plays a crucial role in producing this large anisotropy of interatomic exchange interactions, which persists even for $J_{\rm H}=0$.

\begin{figure}[t!]
\centering
\includegraphics[width=0.96\columnwidth]{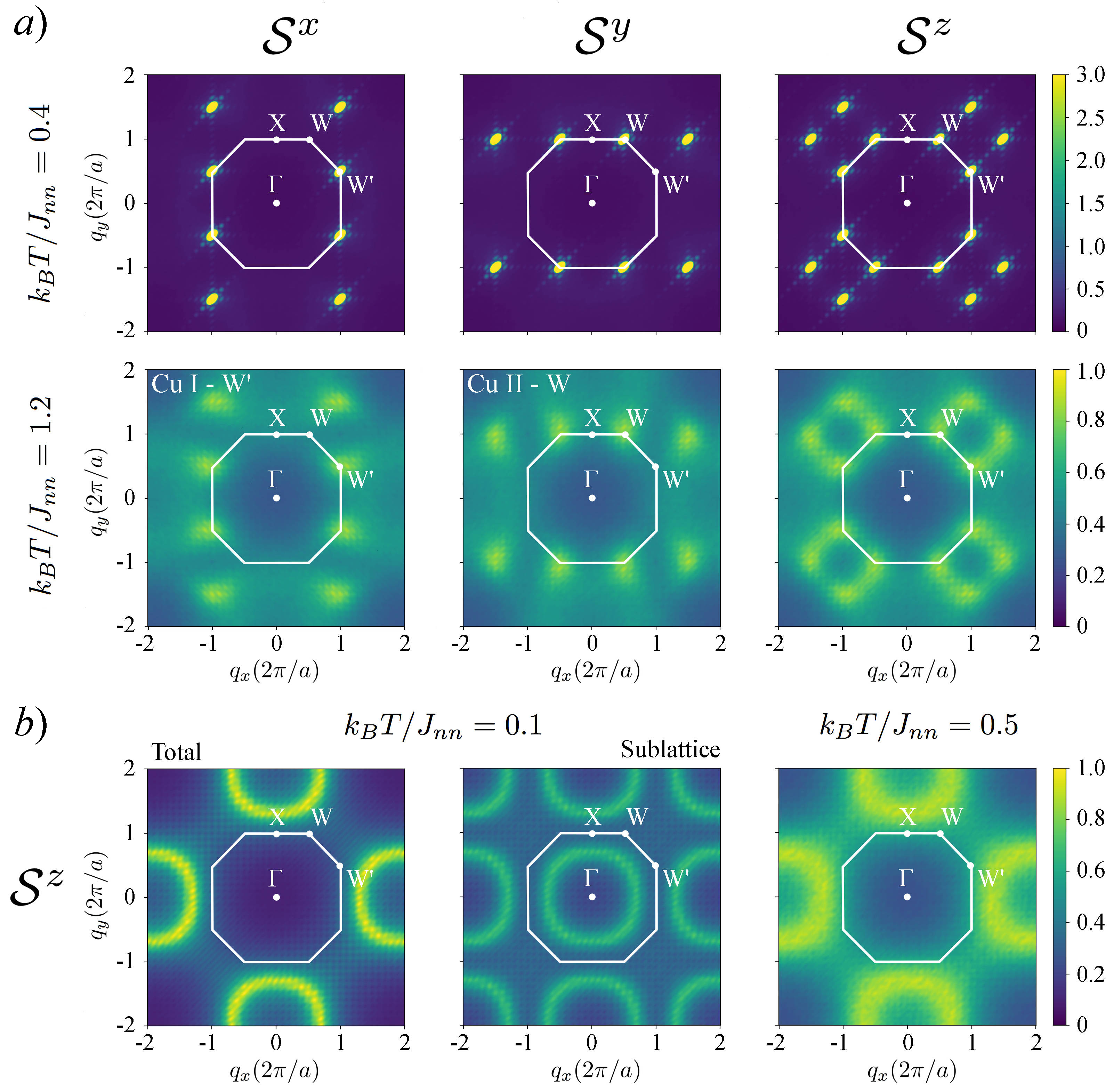}
\caption{Spin structure factors in the $q_{x}q_{y}$ plane of the Brillouin zone as obtained from Monte-Carlo simulations. The results are shown $a)$ for the full $J^{(S)}+J^{(A)}$ model and $b)$~for the $J_{nn}+J_{nnn}$ model without anisotropic exchange calculated for the total structure and one sublattice.}
\label{sscor}
\end{figure}

\emph{Thermodynamic properties}. We perform two independent sets of classical Monte-Carlo simulations based on the conventional Metropolis update algorithm and heat-bath method combined with overrelaxation for the classical spin model~(\ref{eqn:HS})~\cite{supp,montecarlo}. To identify the magnetic order, we calculate the static spin structure factor and specific heat at different temperatures:
\begin{equation}
\begin{aligned}
\mathcal{S}^{\alpha}(\boldsymbol{q})&=\frac{1}{\sqrt{N}}\Big\langle\Big|\sum_{n} e^{-i\boldsymbol{q}\cdot\boldsymbol{R}_{i}}\mathcal{S}_{i}^{\alpha}\Big|\Big\rangle, \\
C_{v}&=\beta^{2}\frac{\langle\mathcal{H}^{2}_{S}\rangle^{\phantom{2}}-{\langle\mathcal{H}^{\phantom{2}}_{S}\rangle^{2}}}{V},
\end{aligned}
\end{equation}
\noindent respectively, where $\langle...\rangle$ stands for the Monte-Carlo averaged configuration, $\beta=1/k_{B}T$, and $V$ is the volume of the supercell. The results are shown in Fig.~\ref{sscor}$a$. Without SO coupling and for isotropic exchange interactions with $J_{nnn}/J_{nn} > 1/8$, a spin-liquid state emerges with the magnetic ground state fluctuating among degenerate spin spirals (Fig.~\ref{sscor}$b$), being in agreement with the analysis by Bergman \textit{et al.}~\cite{Trebst2007}. In turn, anisotropic exchange stabilizes a commensurate spiral state in each sublattice with the $\boldsymbol{q}$-vectors lying in the star of the W point. The result can be understood by considering the energy of the spiral state stabilized by DM interactions~\cite{supp}. Indeed, without loss of generality, one can take $\boldsymbol{\mathcal{S}}_{0}\parallel \boldsymbol{k}_{1}$ and $\boldsymbol{\mathcal{S}}_{i}=\boldsymbol{k}_{1}\cos{(\boldsymbol{R}_{i}\cdot\boldsymbol{q})}+\boldsymbol{k}_{2}\sin{(\boldsymbol{R}_{i}\cdot\boldsymbol{q})}$, where $\boldsymbol{k}_{1}=(0,0,1)$, $\boldsymbol{k}_{2}=(\sin{\phi},-\cos{\phi},0)$, $\boldsymbol{k}_{3}=(\cos{\phi},\sin{\phi},0)$, and $\boldsymbol{q}$ is the propagation vector. Then, summing over all $nnn$ sites and using $\boldsymbol{\mathcal{S}}_{0}\times\boldsymbol{\mathcal{S}}_{i}=\boldsymbol{k}_{3}\sin{(\boldsymbol{R}_{i}\cdot\boldsymbol{q})}$, we have:
\begin{equation}
\begin{aligned}
E_{A}^{\mathrm{I}}&=\frac{d}{2\sqrt{2}}\cos{\phi}\sin{\frac{q_{x}}{2}}\big(\cos{\frac{q_{y}}{2}}-\cos{\frac{q_{z}}{2}}\big) \\
&-\frac{d}{2\sqrt{2}}\sin{\phi}\sin{\frac{q_{y}}{2}}\big(\cos{\frac{q_{x}}{2}}-\cos{\frac{q_{z}}{2}}\big)
\end{aligned}
\end{equation}
\noindent and $E_{A}^{\mathrm{II}} = -E_{A}^{\mathrm{I}}$. In the $q_{x}$-$q_{y}$ plane, the energy is minimised to give two $\boldsymbol{q}$-vectors, $\mathrm{W}=(\pi,2\pi,0)$ and $\mathrm{W'}=(2\pi,\pi,0)$, corresponding to $\tan{\phi}=0$ and $\infty$, respectively. Thus, besides $\mathcal{S}^{z}(\boldsymbol{q})$, which appears in both $\mathrm{W}$ and $\mathrm{W'}$, one can expect finite $\mathcal{S}^{x}(\boldsymbol{q})$ in the $\mathrm{W'}$ point and $\mathcal{S}^{y}(\boldsymbol{q})$ in the $\mathrm{W}$ point, in total agreement with the results shown in Fig.~\ref{sscor}a. Assuming that $J_{nn}$ is not operative at $\boldsymbol{q}=\mathrm{W}$ and $\mathrm{W'}$, where the total interaction with four $nn$ sites vanishes, weak $J_{nnn}$ minimizes the spiral energy at the same $\boldsymbol{q}=\mathrm{W}$ and $\mathrm{W'}$. These conclusions are confirmed by the results of a more rigorous theoretical analysis employing the Luttinger-Tisza method~\cite{luttinger1946,kaplan2007}, which can be summarized as follows~\cite{supp}: (i) DM interactions combined with isotropic exchange alone stabilize the spiral state propagating in each sublattice according to one of the $\boldsymbol{q}$-vectors from the star of the $\mathrm{W}$ point; (ii) The anisotropic symmetric interactions slightly deform the spiral structure.


\par
\begin{figure}[t!]
\centering
\includegraphics[width=0.47\textwidth]{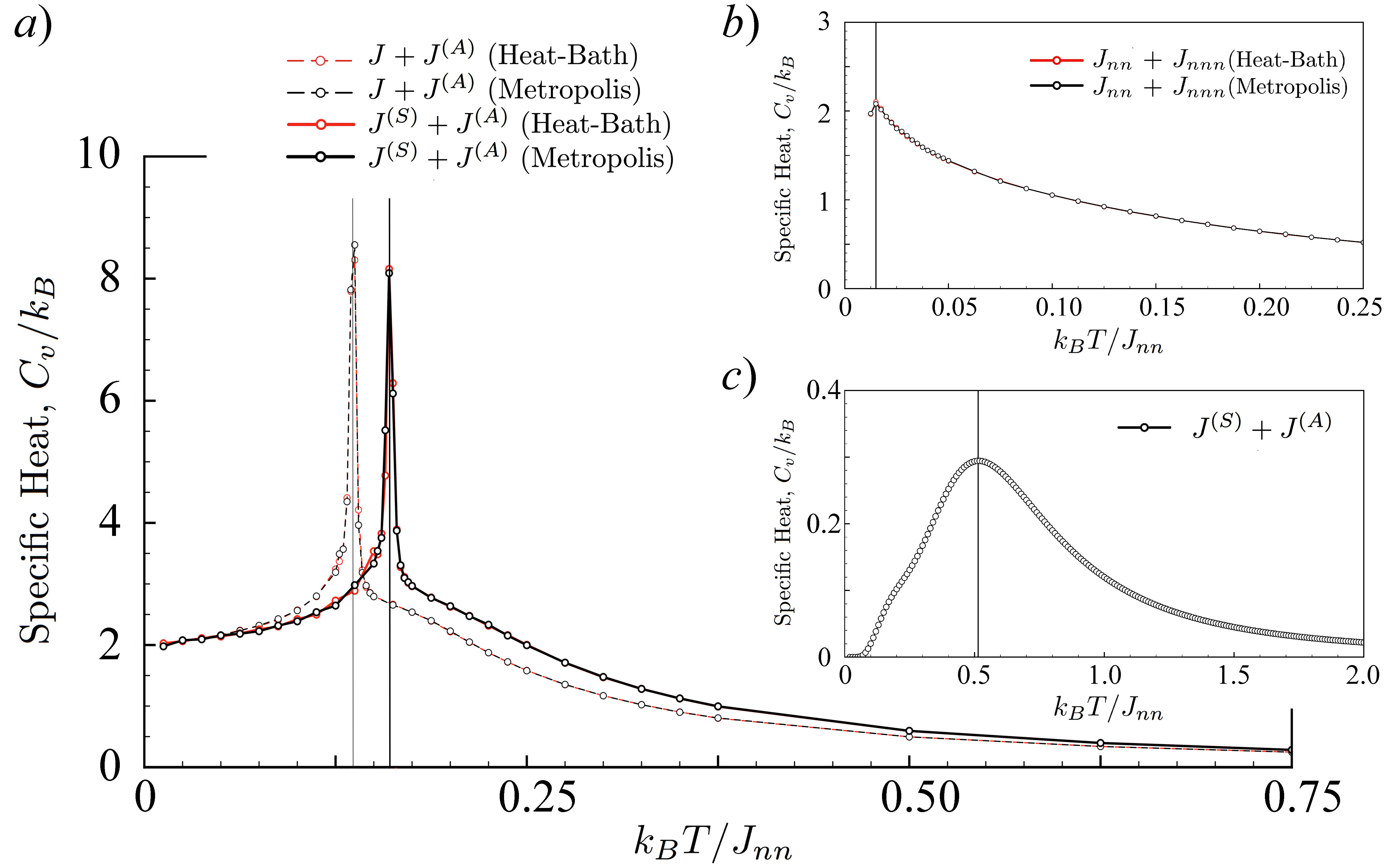}
\caption{ \label{mc-res} $a$)~Specific heat as obtained from classical Monte-Carlo calculations for two models with and without symmetric anisotropic exchange parameters, $J^{(S)}$$+$$J^{(A)}$ and $J^{}$$+$$J^{(A)}$, respectively. $b$)~Specific heat calculated for the classical $J_{nn}$$+$$J_{nnn}$ fully isotropic model. $c$)~Specific heat obtained from exact diagonalization for the quantum $J^{}$$+$$J^{(A)}$ model (with the 4$\times$2$\times$1 periodic supercells containing up to 16 sites). Vertical lines show the corresponding critical temperatures,~$T_{c}$.}
\end{figure}

 The calculated specific heat is shown in Fig.~\ref{mc-res}. According to our results, the system undergoes a magnetic phase transition with the critical temperature $T_{c}\sim 3.2$~K, while without anisotropic exchange the corresponding $T_{c}$ is highly suppressed by magnetic frustration and a certain ordering is stabilized by thermal fluctuations due to an order-by-disorder mechanism at very low temperatures~\cite{Trebst2007}. Finally, to study the effect of quantum fluctuations, we consider the quantum counterpart of Eq.~(\ref{eqn:HS}). Taking into account that the classical ground state is stabilized at the $\mathrm{W}$ point, we assume that the system of $j=1/2$ pseudospins has the same periodicity and solve the quantum model by means of exact diagonalization. Quite expectedly, due to quantum effects the obtained $T_{c}\sim 10.2$ K is higher than in the classical case. Notably, there is an unusual behavior of specific heat below $T_{c}$, where two inflection points are observed. This characteristic is common for $A$-site spinels, where the specific heat departs from a pure $T^{3}$ power law~\cite{Suzuki2007}, and the origin of inflection points is related to the temperature behaviour of magnon spectra~\cite{Bernier2008}.

\emph{Conclusions}. Taking advantage of realistic models derived from first-principles calculations, we have studied the magnetic properties of CuAl$_2$O$_4$. Our results reveal the crucial importance of bond-dependent anisotropic exchange interactions in the physics of Cu spinels, which can be classified as SO Mott insulators. Thus, the range of these materials is not necessarily limited to $5d$ compounds, where SO coupling is large. Similar behavior can be expected in $3d$ oxides with highly symmetric cubic structures. In CuAl$_2$O$_4$, the anisotropic exchange interactions lift the degeneracy of a spiral spin-liquid ground state, anticipated in many $A$-site spinel compounds, and stabilize a commensurate single-$\boldsymbol{q}$ spiral at $T_{c}\sim 10$ K. The experimental verification of our finding requires high quality samples. The currently available ones are characterized by a high level of anti-site disorder, where up to 15\% of Al atoms occupy the Cu $8a$ positions, triggering the spin glass transition~\cite{Nirmala2017}, while in order to observe a genuine magnetic behavior of spinel compounds, the degree of anti-site disorder is typically required to be less than 10\%~\cite{Iakovleva2015,Zaharko2011}. We hope that our theoretical work will stimulate further experimental investigations in this direction.

\emph{Acknowledgments}. We are grateful to C.~H.~Kim, J.-G.~Park, S.~Trebst, and D.~Khomskii for valuable discussions. This work was supported by the Russian Science Foundation through Project No. 17-12-01207.

\end{document}


\title{Supplemental Material:\\ Realization of anisotropic compass model on the diamond lattice of Cu$^{2+}$ in CuAl$_2$O$_4$}

\author{S.~A.~Nikolaev}
\affiliation{International Center for Materials Nanoarchitectonics, National Institute for Materials Science, 1-1 Namiki, Tsukuba, Ibaraki 305-0044, Japan}
\email{saishi@inbox.ru}

\author{I.~V.~Solovyev}
\affiliation{International Center for Materials Nanoarchitectonics, National Institute for Materials Science, 1-1 Namiki, Tsukuba, Ibaraki 305-0044, Japan}
\affiliation{Department of theoretical physics and applied mathematics, Ural Federal University, Mira St. 19, 620002 Yekaterinburg, Russia}
\email{SOLOVYEV.Igor@nims.go.jp}

\author{A.~N.~Ignatenko}
\affiliation{Institute of Metal Physics, S.Kovalevskoy St. 18, 620108  Yekaterinburg, Russia}

\author{V.~Yu.~Irkhin}
\affiliation{Institute of Metal Physics, S.Kovalevskoy St. 18, 620108 Yekaterinburg, Russia}

\author{S.~V.~Streltsov}
\affiliation{Institute of Metal Physics, S.Kovalevskoy St. 18, 620108 Yekaterinburg, Russia}
\affiliation{Department of theoretical physics and applied mathematics, Ural Federal University, Mira St. 19, 620002 Yekaterinburg, Russia}
\email{streltsov@imp.uran.ru}

\date{\today}

\maketitle

\begin{figure}[t!]
\centering
\includegraphics[width=0.8\columnwidth]{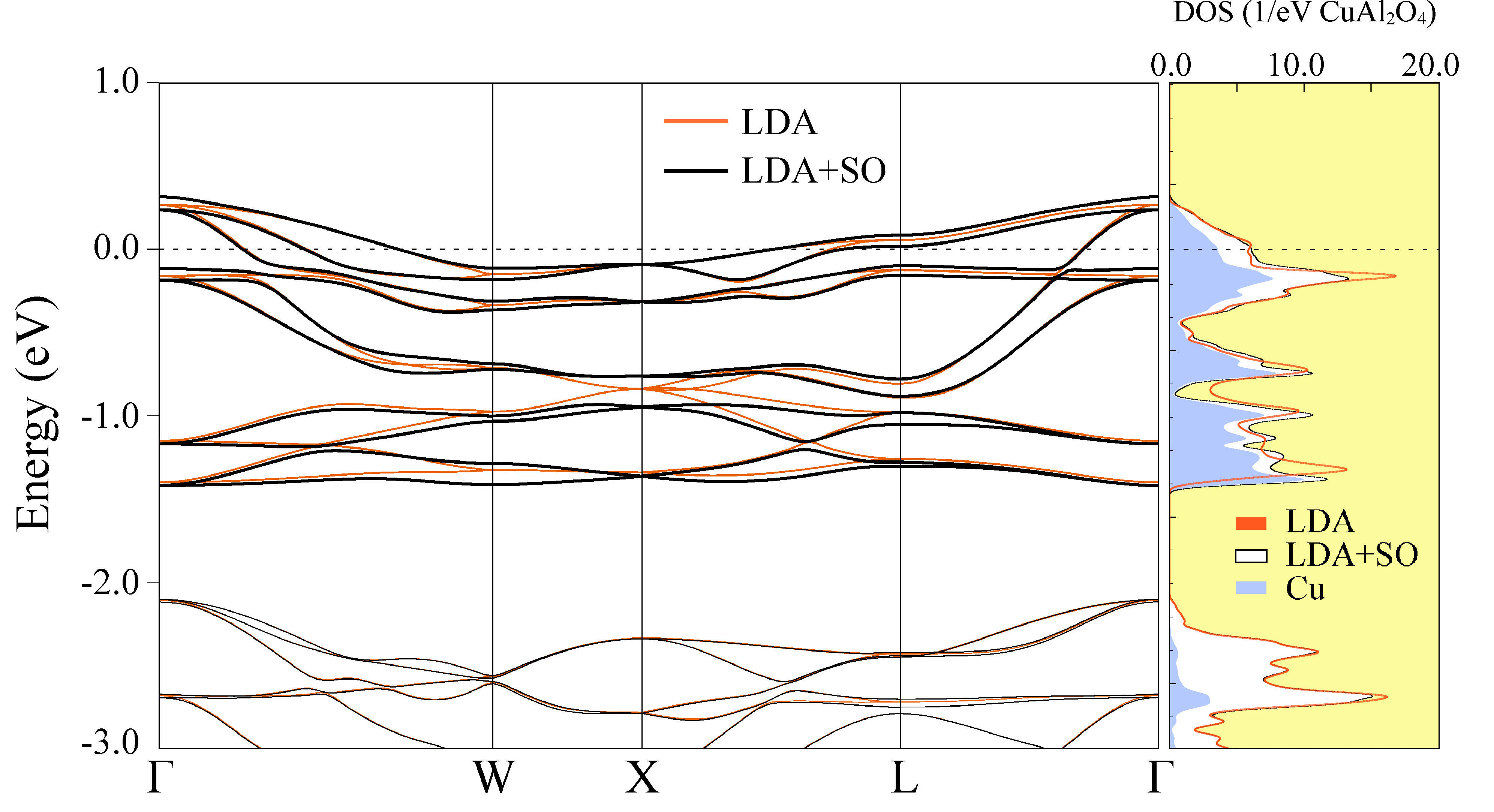}
\caption{Band structure and density of states (DOS) of CuAl$_{2}$O$_{4}$ calculated within local density approximation with and without spin-orbit coupling, LDA+SO and LDA, respectively. The Fermi level is at zero energy.  Highlighted bands represent the basis set of Cu $d$ states used for constructing the low-energy model. Fractional coordinates of the high-symmetry $k$ points are $\mathrm{W}=(\frac{1}{2},\frac{1}{4},\frac{3}{4})$, $\mathrm{X}=(\frac{1}{2},0,\frac{1}{2})$, and $\mathrm{L}=(\frac{1}{2},\frac{1}{2},\frac{1}{2})$.}
\label{fig:band}
\end{figure}

\section{First-principles calculations}
\par Electronic structure calculations of CuAl$_{2}$O$_{4}$ were carried out within the conventional local-density approximation (LDA)~\cite{lda} by using norm-conserving pseudopotentials with and without spin-orbit coupling as implemented in the Quantum ESPRESSO package~\cite{qe}. The plane wave cutoff is set to $\sim$1360 eV, the Brillouin zone is sampled by a $16\times16\times16$ Monkhorst-Pack k-point mesh~\cite{mp}, and the energy convergence criteria is 10$^{-9}$ eV. 

\par The calculated non spin-polarized electronic band structures of CuAl$_{2}$O$_{4}$ with and without spin-orbit coupling are shown in Fig.~\ref{fig:band}. The isolated group of bands near the Fermi level is represented by the Cu $d$ states and is used to construct the effective low-energy Hubbard-type model in the basis of Wannier functions:

\begin{equation}
\hat{\cal{H}}=  \sum_{ij} \sum_{\sigma \sigma'} \sum_{ab}
\left( t_{ij}^{ab}\delta_{\sigma \sigma'}^{} + \Delta^{ab\sigma\sigma'}_{i}\delta_{ij}^{}\right)
\hat{c}^\dagger_{ia\sigma}
\hat{c}^{\phantom{\dagger}}_{jb\sigma'} +  \frac{1}{2}
\sum_{i}  \sum_{\sigma \sigma'} \sum_{abcd} U_{i}^{abcd}
\hat{c}^\dagger_{i a \sigma} \hat{c}^\dagger_{i c \sigma'}
\hat{c}^{\phantom{\dagger}}_{i b \sigma}
\hat{c}^{\phantom{\dagger}}_{i d \sigma'},
\label{eqn.ManyBodyH}
\end{equation}

\noindent where $\hat{c}^{\dagger}_{ia\sigma}$ is the creation operator of an electron with spin $\sigma$ at site $i$ and orbital $a$. The one-electron part of Eq.~(\ref{eqn.ManyBodyH}) is given by hopping parameters $\hat{t}$ and SO coupling $\hat{\Delta}$ (including 10$Dq$ splitting between $e_{g}$ and $t_{2g}$ states), and the second term is the on-site screened Coulomb interaction. To calculate model parameters, we employ the maximally localized Wannier functions as implemented in the Wannier90 package~\cite{wan90}, and the on-site Coulomb parameters are calculated in the Wannier functions basis within the constrained random-phase approximation~\cite{rpa1, rpa2}. The calculated crystal-field splitting between the low-lying $e_{g}$ and high-lying $t_{2g}$ states is $\mu_{e_{g}}-\mu_{t_{2g}}=-0.82$ eV. The latter states are further split by spin-orbit coupling (the corresponding spin-orbit coupling constant is $\xi=0.122$ eV) into the low-lying $j=3/2$ and high-lying $j=1/2$ states. The on-site part of the Hamiltonian can be written as:
$$
\hat{\Delta}_{i}=\left(
\begin{array}{ccccc}
\mu_{t_{2g}} & 0 & 0 & 0 & 0 \\
0 & \mu_{t_{2g}} & 0 & 0 & 0 \\
0 & 0 & \mu_{e_{g}} & 0 & 0 \\
0 & 0 & 0 & \mu_{t_{2g}} & 0 \\
0 & 0 & 0 & 0 & \mu_{e_{g}} \\
\end{array} \right)\hat{\sigma}_{0} +
\frac{\xi}{2}\left(
\begin{array}{ccccc}
0 & \phantom{-}\mathrm{i}\hat{\sigma}_{y} & 0 & -\mathrm{i}\hat{\sigma}_{x} & \phantom{-}2\mathrm{i}\hat{\sigma}_{z} \\
-\mathrm{i}\hat{\sigma}_{y} & 0 & -\sqrt{3}\mathrm{i}\hat{\sigma}_{x} & \phantom{-}\mathrm{i}\hat{\sigma}_{z} & -\mathrm{i}\hat{\sigma}_{x} \\
0 & \phantom{-}\sqrt{3}\mathrm{i}\hat{\sigma}_{x} & 0 & -\sqrt{3}\mathrm{i}\hat{\sigma}_{y} & 0 \\
\phantom{-}\mathrm{i}\hat{\sigma}_{x} &  -\mathrm{i}\hat{\sigma}_{z} & \phantom{-}\sqrt{3}\mathrm{i}\hat{\sigma}_{y} & 0 & -\mathrm{i}\hat{\sigma}_{y} \\
 -2\mathrm{i}\hat{\sigma}_{z} & \phantom{-}\mathrm{i}\hat{\sigma}_{x} & 0 & \phantom{-}\mathrm{i}\hat{\sigma}_{y} & 0 \\
\end{array} \right), 
$$
\noindent where $\hat{\boldsymbol{\sigma}}$ is the vector of Pauli matrices, $\hat{\sigma}_{0}$ is the unity matrix, and orbital indices are in the order $xy$, $yz$, $z^{2}$, $zx$, and $x^{2}-y^{2}$. The rest of the model parameters are given in Tables~II and III.

\begin{table*}[]
\caption{Crystal structure parameters and Wyckoff positions of CuAl$_{2}$O$_{4}$ as obtained from high-resolution neutron and x-ray powder diffraction at 40~K~\cite{nirmala}.}
\begin{center}
\begin{tabular}{c|cc}
\hline
\hline
\multicolumn{3}{c}{$Fd$-$3m$ origin choice 1, $a=8.06983$ \AA}\\
\hline
Cu & 8a & $(0,0,0)$ \\
Al  & 16d & $(\frac{5}{8},\frac{5}{8},\frac{5}{8})$ \\
O & 32e & $(0.38645,0.38645,0.38645)$ \\
\hline
\hline
\end{tabular}
\end{center}
\end{table*}

\section{$j=1/2$ pseudospin model}
\par The low-energy electronic model, Eq.~(\ref{eqn.ManyBodyH}), is mapped onto the spin model of $j=1/2$ pseudospin variables:

\begin{equation}
\mathcal{H}_{S}=\sum\limits_{i>j}\boldsymbol{\mathcal{S}}_{i}\tensor{J}_{ij}\boldsymbol{\mathcal{S}}_{j}+\sum\limits_{i}\boldsymbol{\mathcal{S}}_{i}\tensor{g}\boldsymbol{H},
\label{eq:spin}
\end{equation}

\noindent where $\tensor{J}_{ij}$ is the $3\times3$ tensor of exchange interactions, $\tensor{g}$ is the $g$-factor tensor for the interaction with an external magnetic field $\boldsymbol{H}$, and $\tensor{J}_{ij}$ contains both symmetric $(S)$ and antisymmetric $(A)$ parts, $\tensor{J}_{ij}=\tensor{J}_{ij}^{(S)}+\tensor{J}_{ij}^{(A)}$, where the latter is expressed in terms Dzyaloshinskii-Mroiya interactions, $\boldsymbol{\mathcal{S}}_{i}\tensor{J}_{ij}^{(A)}\boldsymbol{\mathcal{S}}_{j}=\boldsymbol{d}_{ij}\cdot(\boldsymbol{\mathcal{S}}_{i}\times\boldsymbol{\mathcal{S}}_{j})$. As expected for the undistorted cubic structure, the $g$-tensor is isotropic: $g^{\alpha\beta}=g\delta^{\alpha\beta}$, where $g=2$ (the spin and orbital parts are $2/3$ and $4/3$, respectively).

\par  The corresponding model parameters can be derived by applying superexchange theory, where the kinetic energy gain acquired by an unoccupied $j=1/2$ electron at site $i$ in the process of virtual hoppings into the subspace of occupied states at site $j$ (and vice versa) is calculated to second order in hopping processes starting from the atomic limit~\cite{Anderson,solse}:

\begin{equation}
{\cal T}_{ij} = - \frac{1}{S^{2}}\left\langle {\cal G}_{ij}
\left| \hat{t}_{ij} \left( \sum_M \frac{{\hat{\mathscr{P}}}_{j}|{j} M
\rangle \langle {j} M|
{\hat{\mathscr{P}}}_{j}}{E_{{j}M}} \right) \hat{t}_{ji} +
(i \leftrightarrow {j}) \right| {\cal G}_{ij} \right\rangle,
\end{equation}

\noindent where $S=1/2$ is the pseudospin quantum number. Here, ${\cal G}_{ij}$ is the ground-state wavefunction in the atomic limit, constructed in the form of Slater determinants from the high-lying $j=1/2$ Kramer's doublets at sites $i$ and $j$, obtained by diagonalizing the on-site part $\hat{\Delta}$ of Eq.~(\ref{eqn.ManyBodyH}) including spin-orbit coupling and crystal field; $E_{{j}M}$ and $|{j}M \rangle$ stand for the true eigenvalues and eigenvectors of the excited two-electron configurations at site $j$; $\hat{\mathscr{P}}_{j}$ is a projector operator, which enforces the Pauli principle and suppresses any hopping processes into the subspace of unoccupied $j=1/2$ states. The corresponding exchange interactions calculated from the five- and three-orbital electronic models are given in Table~IV.
\par The quantum $j=1/2$ pseudospin model was solved by means of exact diagonalization~\cite{lancsoz} on a $4\times2\times1$ face-centered cubic supercell containing up to 16 pseudospins with periodic boundary conditions. In these calculations, the symmetric anisotropic exchange interactions were neglected on account of their relative smallness compared to the antisymmetric exchange. The classical counterpart of Eq.~(\ref{eq:spin}) is simulated within the conventional Metropolis algorithm and heat-bath method combined with overrelaxation~\cite{montecarlo}. In these calculations, a $20\times20\times20$ face-centered cubic supercell with periodic boundary conditions including $N=1.6\times10^{4}$  classical spins was used, a single run contained up to $2\times10^{6}$ Monte Carlo steps, and the system was gradually cooled down from higher temperatures.

\section{Spin spiral state: analytical approach}
\par The pseudospin model, Eq.~(\ref{eq:spin}), can be solved analytically in the form of a single-$\boldsymbol{q}$ spin spiral,  $
\boldsymbol{\mathcal{S}}_{i}=\boldsymbol{k}_{1}\cos{(\boldsymbol{R}_{i}\cdot\boldsymbol{q})}\sin{\psi}+\boldsymbol{k}_{2}\sin{(\boldsymbol{R}_{i}\cdot\boldsymbol{q})}\sin{\psi} + \boldsymbol{k}_{3}\cos{\psi}$, where the magnetic moment sweeps out a cone of an opening angle $\psi$, and  $\boldsymbol{k}_{i}$'s form a set of three orthonormal vectors, chosen as $\boldsymbol{k}_{1}=(\cos{\phi}\sin{\theta},\sin{\phi}\sin{\theta},\cos{\theta})$, $\boldsymbol{k}_{2}=(\sin{\phi},-\cos{\phi},0)$, and $\boldsymbol{k}_{3}=(\cos{\phi}\cos{\theta},\sin{\phi}\cos{\theta},-\sin{\theta})$ in terms of spherical angles, $\phi$ and $\theta$. Taking into account that (i) $|\boldsymbol{d}|>J_{nnn}$ for each fcc sublattice, and (ii) even though $J_{nn}>J_{nnn}$, $J_{nn}$ is assumed to give no contribution due to the tetrahedral surrounding, we can consider a single fcc sublattice with the antisymmetric exchange. Then, since the Dzyaloshinskii-Moriya interaction favors a perpendicular alignment of two neighboring spins in the same fcc sublattice (global next nearest neighbors), we have $\psi=\pi/2$. Starting from an arbitrary site $0$, so that $\boldsymbol{\mathcal{S}}_{0}=\boldsymbol{k}_{1}$ and $\boldsymbol{\mathcal{S}}_{i}=\boldsymbol{k}_{1}\cos{(\boldsymbol{R}_{i}\cdot\boldsymbol{q})}+\boldsymbol{k}_{2}\sin{(\boldsymbol{R}_{i}\cdot\boldsymbol{q})}$, summing over all nearest neighbors within one subblatice, and using $\boldsymbol{\mathcal{S}}_{0}\times\boldsymbol{\mathcal{S}}_{i}=\boldsymbol{k}_{3}\sin{(\boldsymbol{R}_{i}\cdot\boldsymbol{q})}$, the corresponding energy is written as:

\begin{equation}
\begin{aligned}
E^{\mathrm{I}}_{A}=\frac{S^{2}}{2}\sum\limits_{i}\boldsymbol{d}_{0i}\cdot(\boldsymbol{\mathcal{S}}_{0}\times\boldsymbol{\mathcal{S}}_{i})&=\frac{d}{2\sqrt{2}}\cos{\phi}\cos{\theta}\sin{\frac{q_{x}}{2}}(\cos{\frac{q_{y}}{2}}-\cos{\frac{q_{z}}{2}})-\frac{d}{2\sqrt{2}}\sin{\phi}\cos{\theta}\sin{\frac{q_{y}}{2}}(\cos{\frac{q_{x}}{2}}-\cos{\frac{q_{z}}{2}}) \\
&+\frac{d}{2\sqrt{2}}\sin{\theta}\sin{\frac{q_{z}}{2}}(\cos{\frac{q_{y}}{2}}-\cos{\frac{q_{x}}{2}}).
\end{aligned}
\end{equation}

\noindent Note that $E_{A}^{\mathrm{II}}=-E_{A}^{\mathrm{I}}$. This energy is minimised to give the following $\boldsymbol{q}$ vectors:

\begin{equation}
\begin{array}{ll}
\boldsymbol{q}=(\pi,\pm2\pi,0): & \qquad \theta=0,\,\,\phi=0\qquad (\theta=\pi,\,\,\phi=\pi),\\
\boldsymbol{q}=(-\pi,\pm2\pi,0):& \qquad \theta=0,\,\,\phi=\pi \qquad (\theta=\pi,\,\,\phi=0),\\
\boldsymbol{q}=(\pm2\pi,\pi,0):& \qquad \theta=0,\,\,\phi=-\frac{\pi}{2} \qquad (\theta=\pi,\,\,\phi=\frac{\pi}{2}),\\
\boldsymbol{q}=(\pm2\pi,-\pi,0):& \qquad \theta=0,\,\,\phi=\frac{\pi}{2}, \qquad (\theta=\pi,\,\,\phi=-\frac{\pi}{2}),\\
\boldsymbol{q}=(\pi,0,\pm2\pi):& \qquad \theta=0,\,\,\phi=\pi\qquad (\theta=\pi,\,\,\phi=0),\\
\boldsymbol{q}=(-\pi,0,\pm2\pi):& \qquad \theta=0,\,\,\phi=0 \qquad (\theta=\pi,\,\,\phi=\pi),\\
\boldsymbol{q}=(0,\pi,\pm2\pi):& \qquad \theta=0,\,\,\phi=\frac{\pi}{2} \qquad (\theta=\pi,\,\,\phi=-\frac{\pi}{2}),\\
\boldsymbol{q}=(0,-\pi,\pm2\pi):& \qquad \theta=0,\,\,\phi=-\frac{\pi}{2} \qquad (\theta=\pi,\,\,\phi=\frac{\pi}{2}),\\
\boldsymbol{q}=(\pm2\pi,0,\pi):& \qquad \theta=-\frac{\pi}{2}\qquad(\forall\phi),\\
\boldsymbol{q}=(\pm2\pi,0,-\pi):& \qquad \theta=\frac{\pi}{2}\qquad(\forall\phi),\\
\boldsymbol{q}=(0,\pm2\pi,\pi):& \qquad \theta=\frac{\pi}{2}\qquad(\forall\phi),\\
\boldsymbol{q}=(0,\pm2\pi,-\pi):& \qquad \theta=-\frac{\pi}{2}\qquad(\forall\phi),\\
\end{array}
\end{equation}

\noindent which span the star of the W point.

\section{Luttinger-Tisza method}

\subsection{Classical ground state}

\par In the quasi-momentum representation, Eq.~(\ref{eq:spin}) at zero magnetic field reads:

\begin{equation}
\label{eqn_Hamilt_S}
\mathcal{H}_{S}=\frac{1}{2}\sum_{\boldsymbol{q}}\boldsymbol{\mathcal{S}}^{\dagger}(\boldsymbol{q})
\tensor{J}(\boldsymbol{q})\boldsymbol{\mathcal{S}}(\boldsymbol{q}) 
\end{equation}

\noindent  with

\begin{equation}
\label{eqn_J_block}
\tensor{J}(\boldsymbol{q}) =
\begin{pmatrix}
\tensor{J}^{\mathrm{I,I}}(\boldsymbol{q})  & \tensor{J}^{\mathrm{I,II}}(\boldsymbol{q}) \\
\tensor{J}^{\mathrm{II,I}}(\boldsymbol{q}) & \tensor{J}^{\mathrm{II,II}}(\boldsymbol{q})
\end{pmatrix}
\qquad \mathrm{and} \qquad
\boldsymbol{\mathcal{S}}^{\dagger}(\boldsymbol{q})=
\begin{pmatrix}
\boldsymbol{\mathcal{S}}^{\mathrm{I}*}(\boldsymbol{q}) &
\boldsymbol{\mathcal{S}}^{\mathrm{II}*}(\boldsymbol{q})
\end{pmatrix} ,
\end{equation}

\noindent where

\begin{equation}
\label{eqn_Fourier_S}
\boldsymbol{\mathcal{S}}^{A}(\boldsymbol{q})=\frac{1}{\sqrt{N}}\sum_{n\in A} e^{-i\boldsymbol{q}\cdot\boldsymbol{R}^{A}_{n}}\boldsymbol{\mathcal{S}}^{A}_{n}, \qquad \tensor{J}^{A,A^{\prime}}(\boldsymbol{q})=\sum_{\substack{n^{\prime}\in A'\\n=0\in A}} \tensor{J}(\boldsymbol{R}_{0}^{A}-\boldsymbol{R}_{n^{\prime}}^{A^{\prime}})\, e^{-\mathrm{i}\boldsymbol{q}\cdot(\boldsymbol{R}_{0}^{A}-\boldsymbol{R}_{n'}^{A'})}.
\end{equation}

\noindent Here, $A=\mathrm{I}$ and $\mathrm{II}$ is  the sublattice index, $N$ is the number of lattice cells, summation in $\tensor{J}^{A,A^{\prime}}(\boldsymbol{q})$ is performed over sites $n'$ in sublattice $A^{\prime}$, and $n=0$ is some initial site in sublattice $A$. For nearest neighbors, the exchange interactions are symmetric, $J^{\alpha\beta}(\boldsymbol{R}_{i}^{\mathrm{I}}-\boldsymbol{R}_{j}^{\mathrm{II}})\equiv J_{ij}=J_{nn}\delta^{\alpha\beta}+\Delta J_{nn}(e_{ji}^{\alpha}e_{ij}^{\beta}-\frac{1}{3}\delta^{\alpha\beta})$, where $J_{nn}$ and $\Delta J_{nn}$ are the isotropic and anisotropic components, respectively, and $\boldsymbol{e}_{ji}$ is the unit vector along the bond connecting two sites, and we have:

\begin{equation}
J^{\mathrm{I,II}\,\alpha\beta}(\boldsymbol{q})=4\left(J_{nn} \Lambda_{0}(\boldsymbol{q})\,\delta^{\alpha\beta} - \frac{\Delta J_{nn}}{3} \Lambda^{\alpha\beta}(\boldsymbol{q})\right),
\end{equation}

\noindent where $\Lambda_{0}(\boldsymbol{q})=a(q_x)a(q_y)a(q_z)+ \mathrm{i}\bar{a}(q_x)\bar{a}(q_y)\bar{a}(q_z)$ with $a(x)=\cos(x/4)$, $\bar{a}(x)=\sin(x/4)$; the only nonzero elements of $\tensor{\Lambda}(\boldsymbol{q})$ are the ones with $\alpha\neq\beta$ obtained from $\Lambda_{0}(\boldsymbol{q})$ by interchanging $a(q_{\alpha}) \leftrightarrow \bar{a}(q_{\alpha})$ and $a(q_{\beta}) \leftrightarrow \bar{a}(q_{\beta})$, e.g. $\Lambda^{xy}(\boldsymbol{q}) = \bar{a}(q_x)\bar{a}(q_y)a(q_z) + \mathrm{i}a(q_x)a(q_y)\bar{a}(q_z)$. For next-nearest neighbors, the exchange interactions $\tensor{J}(\boldsymbol{R}^{\mathrm{I}}_{i} - \boldsymbol{R}^{\mathrm{I}}_{j})$ have both symmetric and antisymmetric components, $\tensor{J}_{ij}^{}=\tensor{J}_{ij}^{(S)} + \tensor{J}_{ij}^{(A)}$, where $J_{ij}^{(S)\alpha\beta}=J_{nnn}\delta^{\alpha\beta}+\Delta J_{nnn}(e_{ji}^{\alpha}e_{ji}^{\alpha}-\frac{1}{3})\delta^{\alpha\beta}+\Delta J_{nnn}'(e_{ji}^{\alpha}e_{ji}^{\beta}-\frac{1}{2}\delta^{\alpha\beta})$ and $\tensor{J}^{(A)}_{ij}$ is given by the corresponding Dzyaloshinskii-Moriya vectors as $\boldsymbol{d}_{ij}^{\mathrm{I}(\mathrm{II})}=d[\boldsymbol{n}_{ij}^{\mathrm{I}}\times\boldsymbol{e}_{ji}]$ with $\boldsymbol{n}^{\mathrm{I}}=-\boldsymbol{n}^{\mathrm{II}}=(\mp1,0,0)$ for either $\boldsymbol{e}_{ji} = (0,\frac{1}{\sqrt{2}},\pm \frac{1}{\sqrt{2}})$ or $\boldsymbol{e}_{ji}=(0,-\frac{1}{\sqrt{2}},\mp\frac{1}{\sqrt{2}})$ (for the rest of the bonds one can apply cyclic coordinate permutations), and we have:

\begin{equation}
J^{\mathrm{I,I}\,\alpha\beta}(\boldsymbol{q})=4\big[P^{\alpha}(\boldsymbol{q})\delta^{\alpha\beta}-T^{\alpha\beta}(\boldsymbol{q})\big],
\end{equation}

\noindent where 

\begin{equation}
P^{x}(\boldsymbol{q})= P_{1}b(q_{y})b(q_{z})+P_{2} [b(q_{x})b(q_{z})+b(q_{x})b(q_{y})]
\label{eqn_P}
\end{equation}

\noindent with

\begin{equation}
\begin{gathered}
P_{1}=J_{nnn}-\frac{\Delta J_{nnn}}{3}-\frac{\Delta J_{nnn}^{\prime}}{2},\\
P_{2}=J_{nnn}+\frac{\Delta J_{nnn}}{6},
\end{gathered}
\end{equation}

\noindent $P^{y}(\boldsymbol{q})$ and $P^{y}(\boldsymbol{q})$ are obtained from Eq.~(\ref{eqn_P}) by cyclic permutations of $q_x$, $q_y$, and $q_z$; matrix elements of $\tensor{T}(\boldsymbol{q})$ are nonzero if $\alpha\neq\beta$:

\begin{equation}
T^{\alpha\beta}(\boldsymbol{q})=\frac{\Delta J^{\prime}_{nnn}}{2}\bar{b}(q_{\alpha})\bar{b}(q_{\beta}) + \frac{\mathrm{i}d}{\sqrt{2}}\sum_{\gamma}\varepsilon_{\alpha\beta\gamma} (b(q_{\alpha})-b(q_{\beta}))\bar{b}(q_{\gamma}),
\end{equation}

\noindent where $b(x)=\cos(x/2)$, $\bar{b}(x)=\sin(x/2)$ and $\varepsilon_{\alpha\beta\gamma}$ is the Levi-Civita symbol. Note that $\tensor{J}(\boldsymbol{q})$ is hermitian, and due to inversion symmetry between two sublattices $\tensor{J}^{\mathrm{II,I}}(\boldsymbol{q})=[\tensor{J}^{\mathrm{I,II}}(\boldsymbol{q})]^{*}$ and $\tensor{J}^{\mathrm{II,II}}(\boldsymbol{q})=[\tensor{J}^{\mathrm{I,I}}(\boldsymbol{q})]^{*}$. .

\par To find the classical ground state of  Eq.~(\ref{eqn_Hamilt_S}), we use the Luttinger-Tisza method~\cite{Luttinger_Tisza, Kaplan_Menyuk}. The variables $\boldsymbol{\mathcal{S}}(\boldsymbol{q})$ are not independent and obey the so-called ``strong'' constraints fixing the length of each pseudospin vector, $(\boldsymbol{\mathcal{S}}^{A}_{n})^{2}=S^2$ for all $n\in A$. However, one can impose a much weaker constraint for $\boldsymbol{\mathcal{S}}(\boldsymbol{q})$ to satisfy:

\begin{equation}
\label{eqn_weak_constr}
\sum_{\boldsymbol{q}}\left(|\boldsymbol{\mathcal{S}}^{\mathrm{I}}(\boldsymbol{q})|^{2}+|\boldsymbol{\mathcal{S}}^{\mathrm{II}}(\boldsymbol{q})|^{2}\right) = \sum_{A,n\in A}\big(\boldsymbol{\mathcal{S}}^{A}_{n}\big)^{2}=2NS^2.
\end{equation}

\noindent The Luttinger-Tisza method \cite{Luttinger_Tisza} is based on the observation that every solution to the energy optimization problem (EOP) under the ``weak" constraint which turns out to satisfy ``strong" constraints solves the full EOP too. Having imposed the ``weak" constraint, the EOP of Eq.~(\ref{eqn_weak_constr}) is solved by searching the eigenvectors with minimal eigenvalues of the direct sum $\bigoplus_{\boldsymbol{k}}\tensor{J}(\boldsymbol{k})$ over the whole Brillouin zone. Let the quasi-momentum $\boldsymbol{Q}$ correspond to the global minimum of the smallest eigenvalue of $\tensor{J}(\boldsymbol{q})$, and $\boldsymbol{s}=(\boldsymbol{s}^{\mathrm{I}}\,\,\boldsymbol{s}^{\mathrm{II}})$ be the corresponding eigenvector of $\tensor{J}(\boldsymbol{Q})$. Then, the solution of the EOP under the ``weak" constraint (i.e. the trial classical ground state configuration) is a simple spiral $\boldsymbol{\mathcal{S}}^{A}_{n}=S\,\boldsymbol{m}^{A}_{n}$ with:

\begin{equation}
\label{eqn_simple_spiral}
\begin{pmatrix}
\boldsymbol{m}^{\mathrm{I}}_{n}\\
\boldsymbol{m}^{\mathrm{II}}_{n}
\end{pmatrix}
=
\begin{pmatrix}
\boldsymbol{s}^{\mathrm{I}}e^{\mathrm{i} \boldsymbol{Q}\cdot \boldsymbol{R}^{\mathrm{I}}_{n}}\\
 \boldsymbol{s}^{\mathrm{II}}e^{\mathrm{i} \boldsymbol{Q}\cdot \boldsymbol{R}^{\mathrm{II}}_{n}}
\end{pmatrix}
+ \mathrm{c.c.}\,\,.
\end{equation}

\noindent If the ``strong" constraint is satisfied:

\begin{equation}
\label{eqn_constraint}
1=(\boldsymbol{m}^{A}_{n})^2=2\,(\boldsymbol{s}^{A})^{*}\cdot \boldsymbol{s}^{A} + [(\boldsymbol{s}^{A})^2 e^{\mathrm{i} 2\,\boldsymbol{Q}\cdot \boldsymbol{R}^{A}_{n}}+\mathrm{c.c.}],
\end{equation}

\noindent or, equivalently, $(\boldsymbol{s}^{\mathrm{I}})^2=(\boldsymbol{s}^{\mathrm{II}})^2=0$ and $2\boldsymbol{s}^{\mathrm{I}*}\cdot \boldsymbol{s}^{\mathrm{I}}=2\boldsymbol{s}^{\mathrm{II}*}\cdot \boldsymbol{s}^{\mathrm{II}}=1$, Eq.~(\ref{eqn_simple_spiral}) is a true classical ground state. Otherwise, the Luttinger-Tisza method fails, which, in general, indicates that the determination of classical ground states is a hard nonlinear problem with the macroscopic number of variables. We have performed numerical minimization for the following two cases.

\subsubsection{1. General case including all anisotropies}	
\par In this case, $\boldsymbol{Q}$ is one of the 24 wavevectors obtained by all possible coordinate permutations in the form $(\pm q_1, \pm q_2, 0)$. For the five-orbital (three-orbital) model we find $q_{1}\approx 3.186$ $(3.204)$ and $q_{2}\approx 6.115$ $(6.100)$, and the corresponding $\boldsymbol{Q}$'s are in close vicinity to the vertices of the Brillouin zone, as shown in Fig.~\ref{fig_BZ} with the results of energy minimization for the pseudospin model under the ``weak" constraint, Eq.~(\ref{eqn_weak_constr}). Without loss of generality, consider $\boldsymbol{Q}=(q_1,q_2,0)$, which lies close to the $\mathrm{W}$ point of the Brillouin zone. The corresponding component $\boldsymbol{s}^{\mathrm{I}}$ for the five-orbital (three-orbital) model~is:

\begin{figure}[t]
\includegraphics[width=0.7\columnwidth]{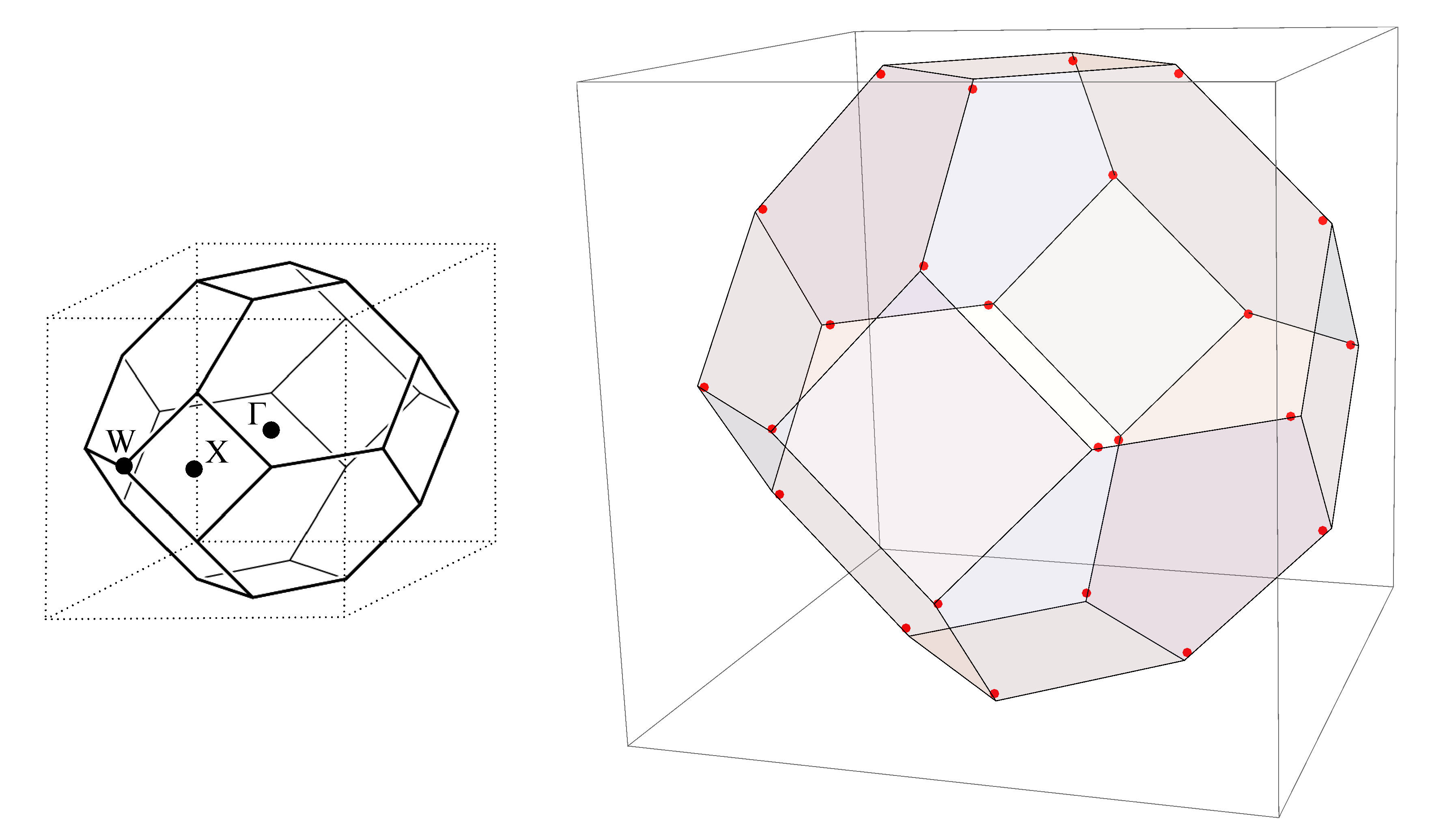}
\caption{\label{fig_BZ} Wavevectors $\boldsymbol{Q}$ (marked by red points) in the fcc Brillouin zone (shown by a truncated octahedron) as obtained after energy minimization of the pseudospin model under the ''weak'' constraint, Eq.~(\ref{eqn_weak_constr}), with the exchange parameters derived from the three-orbital electronic model.}
\end{figure}

\begin{equation}
\boldsymbol{s}^{\mathrm{I}} \approx  (0.041, 0.488, 0.510\,\mathrm{i})\qquad (\boldsymbol{s}^{\mathrm{I}} \approx (0.039, 0.484, 0.514\,\mathrm{i})),
\end{equation}

\noindent and $\boldsymbol{s}^{\mathrm{II}}=\boldsymbol{s}^{\mathrm{I}\,*}$. Since $(\boldsymbol{s}^{\mathrm{I}})^2=(\boldsymbol{s}^{\mathrm{II}})^2\approx -0.020$ $(-0.029)$ is nonzero, the Luttinger-Tisza method fails, and the spiral state, Eq.~(\ref{eqn_simple_spiral}), is not a true ground state. However, since the oscillating terms in Eq.~(\ref{eqn_constraint}), which violate the spin length constraint, are less than three percent compared to the non-oscillating ones, the simple spiral can be considered as a good approximation for the true classical ground state.

\par The resulting proximity of the classical ground state to the simple spiral is unusual. However, as shown below, the simple spiral, Eq.~(\ref{eqn_simple_spiral}), becomes an exact ground state if the symmetric anisotropy $\Delta J_{nn}$ (which is rather small compared to both antisymmetric and remaining symmetric terms) is neglected. \vspace{0.2cm}


\subsubsection{2. All anisotropies except $\Delta J_{nn}$ are included}

\par When the symmetric anisotropy $\Delta J_{nn}$ is discarded, 24 wave vectors $\boldsymbol{Q}$ merge into six nonequivalent wavevectors  $\pm(\pi,2\pi,0)$, $\pm(0,\pi,2\pi)$, and $\pm(2\pi,0,\pi)$ lying at the vertices of the Brillouin zone, as shown in Fig.~\ref{fig_BZ}. In this case, $\tensor{J}^{\mathrm{I,II}}(\boldsymbol{Q})$ and $\tensor{J}^{\mathrm{II,I}}(\boldsymbol{Q})$ vanish, and, as a result,  the corresponding $\tensor{J}(\boldsymbol{Q})$ is block-diagonal, where two sublattices $\mathrm{I}$ and $\mathrm{II}$ are decoupled. Moreover,  owing to $\tensor{J}^{\mathrm{I,I}}(\boldsymbol{Q})=[\tensor{J}^{\mathrm{II,II}}(\boldsymbol{Q})]^{*}$, the smallest  eigenvalue of $\tensor{J}(\boldsymbol{Q})$ (as well as any other) is two-fold degenerate.

\par In the particular case of $\boldsymbol{Q}=(\pi,2\pi,0)$, the corresponding eigenvectors can be written in the form $\widetilde{\boldsymbol{s}}^{T}=\frac{1}{2}(u \boldsymbol{f}\,\, v\boldsymbol{f}^{*})$, where $\boldsymbol{f}=(0,1,\mathrm{i})$ and $u$, $v$ are complex numbers, and the constraint equation, Eq.~(\ref{eqn_constraint}), is satisfied by choosing $u=e^{\mathrm{i}\varphi}$ and $v=e^{\mathrm{i}\psi}$ with some arbitrary phases $\varphi$ and $\psi$. Then, it follows that a simple spiral, Eq.~(\ref{eqn_simple_spiral}) with $\boldsymbol{s}\to\widetilde{\boldsymbol{s}}$, is a true ground state. For the rest of $\boldsymbol{Q}$ obtained by cyclic coordinate permutations, the eigenvector $\widetilde{\boldsymbol{s}}$ has a similar structure with $\boldsymbol{f}$ transformed by the same cyclic permutation. Here,  the transformation $\boldsymbol{Q}\to -\boldsymbol{Q}$ corresponds to $\boldsymbol{f}\to \boldsymbol{f}^{*}$.

\par In fact, the full set of classical ground states is larger since there might be nontrivial linear combinations of simple spirals in the form $\boldsymbol{\mathcal{S}}^{A}_{n}=S\, \boldsymbol{m}^{A}_{n}$ satisfying $(\boldsymbol{m}_{n}^{A})^2 = 1$, where:

\begin{equation}
\label{eqn_linear_combination}
\boldsymbol{m}^{A}_{n}=\sum_{r} \widetilde{\boldsymbol{s}}^{A}_{r} e^{\mathrm{i} \boldsymbol{Q}_r\cdot \boldsymbol{R}^{A}_{n}}+\mathrm{c.c.}
\end{equation}

\noindent with $\boldsymbol{Q}=\boldsymbol{Q}_r$ from the star of the $\mathrm{W}$ point. Here, $\widetilde{\boldsymbol{s}}_{r}^{T}=\frac{1}{2}(u_{r}\boldsymbol{f}_{r}\,\,v_{r}\boldsymbol{f}_{r}^{*})$, where the complex amplitudes $u_{r}$ and $v_{r}$ are to be determined; $r$ runs over the half of the star of the $\mathrm{W}$ point, and the other half obtained by the transformation $\boldsymbol{Q}\to -\boldsymbol{Q}$ is accounted in Eq.~(\ref{eqn_linear_combination}) by complex conjugation. Substituting Eq.~(\ref{eqn_linear_combination}) into $(\boldsymbol{m}_{n}^{A})^2 = 1$ for each sublattice and equating coefficients of the oscillating terms to zero, we have:

\begin{equation}
u_{r}u_{t}=0 \qquad u_{r}u_{t}^{*}=0 \qquad v_{r}v_{t}=0 \qquad v_{r}v_{t}^{*}=0 \qquad \mathrm{for}\,\,r\ne t.
\end{equation}

\noindent Then, it follows that $u_{r}$ and $v_{t}$ are nonzero only for one value of $r=\bar{r}$ and one value of $t=\bar{t}$, respectively. The remaining non-oscillating terms in the constraint equation give $u_{\bar{r}}=e^{\mathrm{i}\varphi}$ and $v_{\bar{t}}=e^{\mathrm{i}\varphi^{\prime}}$ with arbitrary phases $\varphi$ and $\varphi^{\prime}$. Plugging these expressions into Eq. (\ref{eqn_linear_combination}) and denoting $\boldsymbol{Q}=\boldsymbol{Q}_{\bar{r}}$, $\boldsymbol{Q}^{\prime}=\boldsymbol{Q}_{\bar{t}}$, $\boldsymbol{f}=\boldsymbol{f}_{\bar{r}}$, $\boldsymbol{f}^{\prime}=\boldsymbol{f}_{\bar{t}}$, we obtain:

\begin{equation}
\label{eqn_degenerate_GS}
\begin{pmatrix}
\boldsymbol{m}^{\mathrm{I}}_{n}\\
\boldsymbol{m}^{\mathrm{II}}_{n}
\end{pmatrix}
=\frac{1}{2}
\begin{pmatrix}
\boldsymbol{f} e^{\mathrm{i}(\varphi+\boldsymbol{Q}\cdot \boldsymbol{R}^{\mathrm{I}}_{n})}\\
\boldsymbol{f}^{\prime*}e^{\mathrm{i}(\varphi^{\prime}+\boldsymbol{Q}^{\prime}\cdot \boldsymbol{R}^{\mathrm{II}}_{n})}
\end{pmatrix}
+ \mathrm{c.c.}\,.
\end{equation}

\noindent Thus, the most general form of the ground state configuration is represented by two spirals independently propagating in each sublattice $\mathrm{I}$ and $\mathrm{II}$ with the wavevectors $\boldsymbol{Q}$ and $\boldsymbol{Q}^{\prime}$ from the star of the $\mathrm{W}$ point.

\subsection{Structure factor}

\par Finally, let us consider the structure factor when $\Delta J_{nn}=0$:

\begin{equation}
\label{eqn_struct_factor}
\mathcal{S}^{\alpha}(\boldsymbol{q})=\Big\langle\Big|\mathcal{S}^{\mathrm{I}\,\alpha}(\boldsymbol{q}) + \mathcal{S}^{\mathrm{II}\,\alpha}(\boldsymbol{q})\Big|\Big\rangle.
\end{equation}

\noindent In the limit $T\to0$, averaging in Eq.~(\ref{eqn_struct_factor}) is performed over all degenerate ground states $\boldsymbol{\mathcal{S}}^{A}_{n}=S\, \boldsymbol{m}^{A}_{n}$ given by  Eq.~(\ref{eqn_degenerate_GS}). Then, we obtain:

\begin{equation}
\label{eqn_struct_factor_sum}
\mathcal{S}^{\alpha}(\boldsymbol{q})\sim|f_{\alpha} e^{\mathrm{i}\varphi}\delta_{\boldsymbol{q},\boldsymbol{Q}}+f^{*}_{\alpha}e^{-\mathrm{i}\varphi}\delta_{\boldsymbol{q},-\boldsymbol{Q}} + f^{\prime*}_{\alpha}e^{\mathrm{i}\varphi^{\prime}}\delta_{\boldsymbol{q},\boldsymbol{Q}^{\prime}} + f^{\prime}_{\alpha}e^{\mathrm{-i}\varphi^{\prime}}\delta_{\boldsymbol{q},-\boldsymbol{Q}^{\prime}}|.
\end{equation}

\noindent For $\boldsymbol{Q}^{\prime}\not\cong\boldsymbol{Q}$, this expression is reduced to:

\begin{equation}
\label{eqn_Q_nonequiv}
\mathcal{S}^{\alpha}(\boldsymbol{q})\sim|f_{\alpha}|(\delta_{\boldsymbol{q},\boldsymbol{Q}}+\delta_{\boldsymbol{q},-\boldsymbol{Q}})+
|f^{\prime}_{\alpha}|(\delta_{\boldsymbol{q},\boldsymbol{Q}^{\prime}} + \delta_{\boldsymbol{q},-\boldsymbol{Q}^{\prime}}),
\end{equation}

\noindent while for $\boldsymbol{Q}^{\prime}\cong\boldsymbol{Q}$ and, correspondingly, $\boldsymbol{f}^{\prime}=\boldsymbol{f}$, we have:

\begin{equation}
\label{eqn_Q_equiv} 
\mathcal{S}^{\alpha}(\boldsymbol{q})\sim|f_{\alpha}+f^{*}_{\alpha}e^{\mathrm{i}\psi}|(\delta_{\boldsymbol{q},\boldsymbol{Q}}+\delta_{\boldsymbol{q},-\boldsymbol{Q}}),
\end{equation}

\noindent where $\psi=\varphi^{\prime}-\varphi$. Let us consider $\boldsymbol{q}$ lying in the $q_{x}$-$q_{y}$ plane. Then, Eqs.~(\ref{eqn_Q_nonequiv}) and (\ref{eqn_Q_equiv}) are nonzero only when $\boldsymbol{Q}$ (or $\boldsymbol{Q}^{\prime}$) is one of the four wavevectors $\boldsymbol{Q}_{1}=(\pi,2\pi,0)$, $\boldsymbol{Q}_{2}=(2\pi,\pi,0)$, $-\boldsymbol{Q}_{1}$, and $-\boldsymbol{Q}_{2}$. The corresponding vectors $\boldsymbol{f}$ (or $\boldsymbol{f}^{\prime}$) are $\boldsymbol{f}_{1}=(0,1,\mathrm{i})$, $\boldsymbol{f}_{2}=(-\mathrm{i},0,1)$, $\boldsymbol{f}_{1}^{*}$, and $\boldsymbol{f}_{2}^{*}$, respectively. Next, we average Eqs.~(\ref{eqn_Q_nonequiv}) and (\ref{eqn_Q_equiv}) over $\psi$, $\boldsymbol{Q}$, and $\boldsymbol{Q}^{\prime}$. In Eq.~(\ref{eqn_Q_nonequiv}), owing to the correspondence between $\boldsymbol{Q}$ and $\boldsymbol{f}$, only two wavevectors $\pm \boldsymbol{Q}_{2}$ ($\pm \boldsymbol{Q}_{1}$) contribute to the $x$ ($y$) component, whereas all four vectors, $\pm \boldsymbol{Q}_{2}$ and $\pm \boldsymbol{Q}_{1}$, contribute to the $z$ component. All these contributions have the same weight. In Eq.~(\ref{eqn_Q_equiv}), we have similar contributions, but their weights  are proportional to $|\cos(\psi/2)|$ or $|\sin(\psi/2)|$, which equate only after integrating over $\psi$. Thus, the contributions to the structure factor coming from Eqs. (\ref{eqn_Q_nonequiv}) and (\ref{eqn_Q_equiv}) are proportional to each other. Consequently, at $T\to 0$ for $\boldsymbol{q}$ lying in the $q_{x} $-$q_{y}$ plane we have:

\begin{equation}
\begin{gathered}
\mathcal{S}^{x}(\boldsymbol{q})= K(\delta_{\boldsymbol{q},\boldsymbol{Q}_{2}}+\delta_{\boldsymbol{q},-\boldsymbol{Q}_{2}}),\\
\mathcal{S}^{y}(\boldsymbol{q})= K(\delta_{\boldsymbol{q},\boldsymbol{Q}_{1}}+\delta_{\boldsymbol{q},-\boldsymbol{Q}_{1}}),\\
\mathcal{S}^{z}(\boldsymbol{q})=\mathcal{S}^{x}(\boldsymbol{q})+\mathcal{S}^{y}(\boldsymbol{q}),
\end{gathered}
\end{equation}

\noindent where $K=\mathrm{const}$. The obtained pattern of peaks is in perfect agreement with the results shown in Fig.~2(a) of the main text.

\begin{table*}
\caption{Nearest and next-nearest neighbor hopping parameters (in meV) calculated for the five-orbital model of CuAl$_{2}$O$_{4}$. The corresponding hopping parameters for Cu II are $\hat{t}^{\mathrm{II},\mathrm{I}}_{ji}=\hat{t}^{\mathrm{I},\mathrm{II}}_{ij}$ and $\hat{t}^{\mathrm{II},\mathrm{II}}_{ij}=[\hat{t}^{\mathrm{I},\mathrm{I}}_{ij}]^{T}$. Orbital indices are given in the order $xy$, $yz$, $z^{2}$, $zx$, and $x^{2}-y^{2}$. Bonds are in units of the lattice constant, $a$.}
\begin{center}
\begin{tabular}{c|c|c|c}
\hline
\hline
Bond&&Bond& \\
\hline
$(\frac{1}{4},\frac{1}{4},\frac{1}{4})$ &
$
\hat{t}^{\mathrm{I},\mathrm{II}}_{ij}=\left(
\begin{array}{rrrrr}
 -41.3 &-24.1 &-3.3 &-24.1 & 0.0 \\
 -24.1 &-41.3 & 1.7 &-24.1 & -2.9 \\
 -3.3 & 1.7 & 50.5 & 1.7 & 0.0 \\
 -24.1 & -24.1 & 1.7 & -41.3 & 2.9 \\
 0.0 &-2.9 & 0.0 & 2.9 & 50.5 \\
\end{array} \right)
$ &
$(-\frac{1}{4},-\frac{1}{4},\frac{1}{4})$ &
$
\hat{t}^{\mathrm{I},\mathrm{II}}_{ij}=\left(
\begin{array}{rrrrr}
 -41.3 & 24.1 &-3.3 & 24.1 & 0.0 \\
  24.1 &-41.3 &-1.7 &-24.1 & 2.9 \\
 -3.3 &-1.7 & 50.5 &-1.7 & 0.0 \\
  24.1 &-24.1 &-1.7 &-41.3 &-2.9 \\
  0.0 & 2.9 & 0.0 &-2.9 & 50.5 \\
\end{array} \right)
$ \\
&&& \\
$(-\frac{1}{4},\frac{1}{4},-\frac{1}{4})$ &
$
\hat{t}^{\mathrm{I},\mathrm{II}}_{ij}=\left(
\begin{array}{rrrrr}
 -41.3 &-24.1 & 3.3 & 24.1 & 0.0 \\
 -24.1 &-41.3 &-1.7 & 24.1 & 2.9 \\
  3.3 &-1.7 & 50.5 & 1.7 & 0.0 \\
  24.1 & 24.1 & 1.7 &-41.3 & 2.9 \\
 0.0 & 2.9 & 0.0 & 2.9 & 50.5 \\
\end{array} \right)
$ &
$(\frac{1}{4},-\frac{1}{4},-\frac{1}{4})$ &
$
\hat{t}^{\mathrm{I},\mathrm{II}}_{ij}=\left(
\begin{array}{rrrrr}
 -41.3 & 24.1 & 3.3 &-24.1 & 0.0 \\
  24.1& -41.3 & 1.7 & 24.1 &-2.9 \\ 
  3.3 & 1.7 & 50.5 &-1.7 & 0.0 \\
 -24.1 & 24.1 &-1.7 & -41.3 &-2.9 \\
  0.0 &-2.9 &0.0 &-2.9 & 50.5 \\
\end{array} \right)
$ \\
&&& \\
\hline
&&& \\
$(0,\frac{1}{2},\frac{1}{2})$ &
$
\hat{t}^{\mathrm{I},\mathrm{I}}_{ij}=\left(
\begin{array}{rrrrr}
  49.8 & 19.5 & -15.9 & 57.0 & 37.0 \\
 -19.5 &-15.9 & 5.4 &-19.5 &-9.4 \\
  15.9  & 5.4  & 3.8 & 24.1 & 22.2 \\
  57.0 & 19.5 & -24.1 & 49.8 & 32.3 \\
 -37.0 & -9.4 & 22.2 & -32.3 & -21.8 \\
\end{array} \right)
$ &
$(-\frac{1}{2},0,\frac{1}{2})$ &
$
\hat{t}^{\mathrm{I},\mathrm{I}}_{ij}=\left(
\begin{array}{rrrrr}
  49.8 & -57.0 & -15.9 & -19.5 & -37.0 \\
 -57.0 & 49.8 & 24.1 & 19.5 & 32.3 \\
  15.9 & -24.1 & 3.8 & -5.4 & -22.2 \\
  19.5 & -19.5 & -5.4 & -15.9 & -9.4 \\
  37.0 & -32.3 & -22.2 & -9.4 & -21.8 \\
\end{array} \right)
$ \\
&&& \\
$(-\frac{1}{2},\frac{1}{2},0)$ &
$
\hat{t}^{\mathrm{I},\mathrm{I}}_{ij}=\left(
\begin{array}{rrrrr}
 -15.9 & -19.5 & 10.8  & 19.5 & 0.0 \\
  19.5  & 49.8 & -40.0 & -57.0 & -4.7 \\
  10.8  & 40.0 & -34.6 & -40.0 & 0.0 \\
 -19.5 & -57.0 & 40.0 & 49.8 & -4.7 \\
  0.0 & 4.7 & 0.0 & 4.7 & 16.6 \\
\end{array} \right)
$ &
$(\frac{1}{2},0,\frac{1}{2})$ &
$
\hat{t}^{\mathrm{I},\mathrm{I}}_{ij}=\left(
\begin{array}{rrrrr}
  49.8 & 57.0 &-15.9 & 19.5 &-37.0 \\
  57.0 & 49.8 &-24.1 & 19.5 &-32.3 \\
  15.9 & 24.1 & 3.8 & 5.4 &-22.2 \\
 -19.5 &-19.5 & 5.4 &-15.9 & 9.4 \\
  37.0 & 32.3 &-22.2 & 9.4 &-21.8 \\
\end{array} \right)
$ \\
&&&\\
$(0,-\frac{1}{2},\frac{1}{2})$ &
$
\hat{t}^{\mathrm{I},\mathrm{I}}_{ij}=\left(
\begin{array}{rrrrr}
  49.8 &-19.5 &-15.9 &-57.0 & 37.0 \\
  19.5 &-15.9 &-5.4 &-19.5 & 9.4 \\
  15.9 &-5.4 & 3.8 &-24.1 & 22.2 \\
 -57.0 & 19.5 & 24.1 & 49.8 &-32.3 \\
 -37.0 & 9.4 & 22.2 & 32.3 &-21.8 \\
\end{array} \right)
$ &
$(\frac{1}{2},\frac{1}{2},0)$ &
$
\hat{t}^{\mathrm{I},\mathrm{I}}_{ij}=\left(
\begin{array}{rrrrr}
 -15.9 &-19.5 &-10.8 &-19.5 & 0.0 \\
  19.5 & 49.8 & 40.0 & 57.0 & 4.7 \\
 -10.8 &-40.0 &-34.6 &-40.0 & 0.0 \\
  19.5 & 57.0 & 40.0 & 49.8 &-4.7 \\
  0.0 &-4.7 & 0.0 & 4.7 & 16.6 \\
\end{array} \right)
$ \\
&&& \\
$(-\frac{1}{2},-\frac{1}{2},0)$ &
$
\hat{t}^{\mathrm{I},\mathrm{I}}_{ij}=\left(
\begin{array}{rrrrr}
 -15.9 & 19.5 &-10.8 & 19.5 & 0.0 \\
 -19.5 & 49.8 &-40.0 & 57.0 &-4.7 \\
 -10.8 & 40.0 &-34.6 & 40.0 &0.0 \\
 -19.5 & 57.0 &-40.0 & 49.8 & 4.7 \\
  0.0 & 4.7 &0.0 &-4.7 & 16.6 \\
\end{array} \right)
$ &
$(0,\frac{1}{2},-\frac{1}{2})$ &
$
\hat{t}^{\mathrm{I},\mathrm{I}}_{ij}=\left(
\begin{array}{rrrrr}
  49.8 & 19.5 & 15.9 &-57.0 &-37.0 \\
 -19.5 &-15.9 &-5.4 & 19.5 & 9.4 \\
 -15.9 &-5.4 & 3.8 & 24.1 & 22.2 \\
 -57.0 &-19.5 &-24.1 & 49.8 & 32.3 \\
  37.0 & 9.4 & 22.2 &-32.3 &-21.8 \\
\end{array} \right)
$ \\
&&&\\
$(-\frac{1}{2},0,-\frac{1}{2})$ &
$
\hat{t}^{\mathrm{I},\mathrm{I}}_{ij}=\left(
\begin{array}{rrrrr}
  49.8 & 57.0 & 15.9 &-19.5 & 37.0 \\
  57.0 & 49.8 & 24.1 &-19.5 & 32.3 \\
 -15.9 &-24.1 & 3.8 & 5.4 &-22.2 \\
  19.5 & 19.5 & 5.4 &-15.9 & 9.4 \\
 -37.0 &-32.3 &-22.2 & 9.4 &-21.8 \\
\end{array} \right)
$ &
$(\frac{1}{2},-\frac{1}{2},0)$ &
$
\hat{t}^{\mathrm{I},\mathrm{I}}_{ij}=\left(
\begin{array}{rrrrr}
 -15.9 & 19.5 & 10.8 &-19.5 & 0.0 \\ 
 -19.5 & 49.8 & 40.0 &-57.0 & 4.7 \\
  10.8 &-40.0 &-34.6 & 40.0 & 0.0 \\
  19.5 &-57.0 &-40.0 & 49.8 & 4.7 \\
  0.0 &-4.7 & 0.0 &-4.7 & 16.6 \\
\end{array} \right)
$ \\
&&&\\
$(\frac{1}{2},0,-\frac{1}{2})$ &
$
\hat{t}^{\mathrm{I},\mathrm{I}}_{ij}=\left(
\begin{array}{rrrrr}
  49.8 &-57.0 & 15.9 & 19.5 & 37.0 \\
 -57.0 & 49.8 &-24.1 &-19.5 &-32.3 \\
 -15.9 & 24.1 & 3.8 &-5.4 &-22.2 \\
 -19.5 & 19.5 &-5.4 &-15.9 &-9.4 \\
 -37.0 & 32.3 &-22.2 &-9.4 &-21.8 \\
\end{array} \right)
$ &
$(0,-\frac{1}{2},-\frac{1}{2})$ &
$
\hat{t}^{\mathrm{I},\mathrm{I}}_{ij}=\left(
\begin{array}{rrrrr}
  49.8 &-19.5 & 15.9 & 57.0 &-37.0 \\
  19.5 &-15.9 & 5.4 & 19.5 &-9.4 \\
 -15.9 & 5.4 & 3.8 &-24.1 & 22.2 \\
  57.0 &-19.5 & 24.1 & 49.8 &-32.3 \\
  37.0 &-9.4 & 22.2 & 32.3 &-21.8 \\
\end{array} \right)
$ \\
&&&\\
\hline
\hline
\end{tabular}
\end{center}
\end{table*}

\begin{table*}
\caption{Partially screened Coulomb interaction parameters (in eV) calculated for the five-orbital model of CuAl$_{2}$O$_{4}$. Orbital indices are given in the order $xy$, $yz$, $z^{2}$, $zx$, and $x^{2}-y^{2}$.}
\begin{center}
\begin{tabular}{c}
\hline
\hline
\\
$
U^{mm'mm'}=\left(
\begin{array}{ccccc}
   5.3975 &  3.9274 &  3.8091 &  3.9274 &  4.2469 \\
   3.9274 &  5.5066 &  4.5435 &  4.0016 &  3.6880 \\
   3.8091 &  4.5435 &  5.7536 &  4.5435 &  3.5769 \\
   3.9274 &  4.0016 &  4.5435 &  5.5066 &  3.6880 \\
   4.2469 &  3.6880 &  3.5769 &  3.6880 &  4.8093 \\
\end{array} \right)
$ \\
\\
$
U^{mmm'm'}=\left(
\begin{array}{ccccc}
   5.3975 &  0.7636 &  0.8917 &  0.7636 &  0.4275 \\
   0.7636 &  5.5066 &  0.5433 &  0.7608 &  0.7297 \\
   0.8917 &  0.5433 &  5.7536 &  0.5433 &  0.8475 \\ 
   0.7636 &  0.7608 &  0.5433 &  5.5066 &  0.7297 \\
   0.4275 &  0.7297 &  0.8475 &  0.7297 &  4.8093 \\
\end{array} \right)
$ \\
\\
$
U^{mm'm'm}=\left(
\begin{array}{ccccc}
   5.3975 &  0.7636 &  0.8917 &  0.7636 &  0.4275 \\
   0.7636 &  5.5066 &  0.5433 &  0.7608 &  0.7297 \\
   0.8917 &  0.5433 &  5.7536 &  0.5433 &  0.8475 \\
   0.7636 &  0.7608 &  0.5433 &  5.5066 &  0.7297 \\
   0.4275 &  0.7297 &  0.8475 &  0.7297 &  4.8093 \\
\end{array} \right)
$  \\
\\
\hline
\hline
\end{tabular}
\end{center}
\end{table*}

\newpage
\vspace{6.0cm}
\LTcapwidth=\linewidth
\begin{longtable}[t!]{c|c|cc}
\caption{Nearest and next-nearest neighbor exchange interactions (in meV) calculated from the five- and three-orbital electronic models of CuAl$_{2}$O$_{4}$. The corresponding exchange parameters for Cu II are $\tensor{J}^{\mathrm{II},\mathrm{I}}_{ji}=\tensor{J}^{\mathrm{I},\mathrm{II}}_{ij}$ and $\tensor{J}^{\mathrm{II},\mathrm{II}}_{ij}=[\tensor{J}^{\mathrm{I},\mathrm{I}}_{ij}]^{T}$. Bonds are in units of the lattice constant, $a$. \vspace{0.2cm}}
\\
\hline
\hline
\vspace{0.1cm}\\
Bond & Unit vector, $\boldsymbol{e}_{ji}$ & Five-orbital model & Three-orbital model \\
$\phantom{A}$&&&\\
\hline
$\phantom{A}$&&&\\
& & \multicolumn{2}{c}{Nearest neighbors} \\
& & $\tensor{J}^{\mathrm{I},\mathrm{II}}_{ij}$ &  $\tensor{J}^{\mathrm{I},\mathrm{II}}_{ij}$  \\
$\phantom{A}$&&&\\
$(\frac{1}{4},\frac{1}{4},\frac{1}{4})$ & $(\frac{1}{\sqrt{3}},\frac{1}{\sqrt{3}},\frac{1}{\sqrt{3}})$ &
$
\left(
\begin{array}{rrr}
  1.724 &   0.080 &    0.076 \\
  0.080 &   1.724 &    0.076 \\
  0.076 &   0.076 &    1.728 \\
\end{array} \right)
$ &
$
\left(
\begin{array}{rrr}
  2.096 &    0.076 &    0.068 \\
  0.076 &    2.096 &    0.068 \\
  0.068 &    0.068 &    2.104 \\
\end{array} \right)
$
\\
$\phantom{A}$&&&\\
$(-\frac{1}{4},\frac{1}{4},-\frac{1}{4})$ & $(-\frac{1}{\sqrt{3}},\frac{1}{\sqrt{3}},-\frac{1}{\sqrt{3}})$ &
$
\left(
\begin{array}{rrr}
  1.724 &   -0.080 &    0.076 \\
  -0.080 &   1.724 &    -0.076 \\
  0.076 &   -0.076 &    1.728 \\
\end{array} \right)
$ &
$
\left(
\begin{array}{rrr}
  2.096 &    -0.076 &    0.068 \\
  -0.076 &    2.096 &    -0.068 \\
  0.068 &    -0.068 &    2.104 \\
\end{array} \right)
$
\\
$\phantom{A}$&&&\\
$(-\frac{1}{4},-\frac{1}{4},\frac{1}{4})$ & $(-\frac{1}{\sqrt{3}},-\frac{1}{\sqrt{3}},\frac{1}{\sqrt{3}})$ &
$
\left(
\begin{array}{rrr}
  1.724 &   0.080 &    -0.076 \\
  0.080 &   1.724 &    -0.076 \\
  -0.076 &   -0.076 &    1.728 \\
\end{array} \right)
$ &
$
\left(
\begin{array}{rrr}
  2.096 &    0.076 &    -0.068 \\
  0.076 &    2.096 &    -0.068 \\
  -0.068 &    -0.068 &    2.104 \\
\end{array} \right)
$
\\
$\phantom{A}$&&&\\
$(\frac{1}{4},-\frac{1}{4},-\frac{1}{4})$ & $(\frac{1}{\sqrt{3}},-\frac{1}{\sqrt{3}},-\frac{1}{\sqrt{3}})$ &
$
\left(
\begin{array}{rrr}
1.724 &   -0.080 &    -0.076 \\
  -0.080 &   1.724 &    0.076 \\
  -0.076 &   0.076 &    1.728 \\
\end{array} \right)
$ &
$
\left(
\begin{array}{rrr}
  2.096 &    -0.076 &    -0.068 \\
  -0.076 &    2.096 &    0.068 \\
  -0.068 &    0.068 &    2.104 \\
\end{array} \right)
$
\\
$\phantom{A}$&&&\\
\hline
$\phantom{A}$&&&\\
& & \multicolumn{2}{c}{Next-nearest neighbors} \\
&& $\tensor{J}^{\mathrm{I},\mathrm{I}}_{ij}$ & $\tensor{J}^{\mathrm{I},\mathrm{I}}_{ij}$ \\
$\phantom{A}$&&&\\
$(0,\frac{1}{2},\frac{1}{2})$ & $(0,\frac{1}{\sqrt{2}},\frac{1}{\sqrt{2}})$ &
$
\left(
\begin{array}{rrr}
 0.068 &   -0.776 &   -0.776 \\
 0.776 &    0.664 &    0.032 \\
 0.776 &    0.032 &    0.660 \\
\end{array} \right)
$ &
$
\left(
\begin{array}{rrr}
 0.004 &   -0.736 &   -0.736 \\
 0.736 &    0.672 &    0.0 \\
 0.736 &    0.0 &    0.660 \\
\end{array} \right)
$ 
\\ 
$\phantom{A}$&&&\\
$(-\frac{1}{2},0,\frac{1}{2})$ & $(-\frac{1}{\sqrt{2}},0,\frac{1}{\sqrt{2}})$ &
$
\left(
\begin{array}{rrr}
 0.664  & 0.776 & -0.032 \\
-0.776 & 0.068 & 0.776 \\
-0.032 & -0.776 & 0.660 \\
\end{array} \right)
$ &
$
\left(
\begin{array}{rrr}
   0.672 &  0.736 &  0.0 \\
  -0.736 &  0.004 &  0.736 \\
  0.0 & -0.736 & 0.660 \\
\end{array} \right)
$ 
\\ 
$\phantom{A}$&&&\\
$(-\frac{1}{2},\frac{1}{2},0)$ & $(-\frac{1}{\sqrt{2}},\frac{1}{\sqrt{2}},0)$ &
$
\left(
\begin{array}{rrr}
  0.636 &-0.032 & 0.776 \\
 -0.032 & 0.636 &-0.776 \\
 -0.776 & 0.776 & 0.080 \\
\end{array} \right)
$ &
$
\left(
\begin{array}{rrr}
  0.604 & 0.0 & 0.732 \\
  0.0 & 0.604 & -0.732 \\
 -0.732 & 0.732 & 0.0 \\
\end{array} \right)
$ 
\\
$\phantom{A}$&&&\\ 
$(\frac{1}{2},0,\frac{1}{2})$ & $(\frac{1}{\sqrt{2}},0,\frac{1}{\sqrt{2}})$ &
$
\left(
\begin{array}{rrr}
  0.664 &  0.776 & 0.032 \\
 -0.776 & 0.068 & -0.776 \\
  0.032 &  0.776 & 0.660 \\
\end{array} \right)
$ &
$
\left(
\begin{array}{rrr}
   0.668 &   0.736 &    0.0 \\
  -0.736 &    0.004 &   -0.736 \\
   0.0 &    0.736 &   0.660 \\
\end{array} \right)
$
\\
$\phantom{A}$&&&\\
$(0,-\frac{1}{2},\frac{1}{2})$ & $(0,-\frac{1}{\sqrt{2}},\frac{1}{\sqrt{2}})$ &
$
\left(
\begin{array}{rrr}
 0.068 & -0.776 & 0.776 \\
 0.776 &  0.664 & -0.032 \\
-0.776 & -0.032 & 0.660 \\
\end{array} \right)
$ &
$
\left(
\begin{array}{rrr}
  0.016  & -0.736 &    0.736 \\
  0.736 &   0.672 &   0.0 \\
 -0.736 &   0.0 &    0.660 \\
\end{array} \right)
$
\\
$\phantom{A}$&&&\\
$(\frac{1}{2},\frac{1}{2},0)$ & $(\frac{1}{\sqrt{2}},\frac{1}{\sqrt{2}},0)$ &
$
\left(
\begin{array}{rrr}
 0.636 & 0.032 & 0.776 \\
 0.032 & 0.636 & 0.776 \\
-0.776 & -0.776 & 0.080 \\
\end{array} \right)
$ &
$
\left(
\begin{array}{rrr}
  0.604 &  0.0 &    0.732 \\
 0.0 &  0.604 &    0.732\\
 -0.732 & -0.732 &  0.004 \\
\end{array} \right)
$
\\
$\phantom{A}$&&&\\
$(-\frac{1}{2},-\frac{1}{2},0)$ & $(-\frac{1}{\sqrt{2}},-\frac{1}{\sqrt{2}},0)$ &
$
\left(
\begin{array}{rrr}
   0.636 &    0.032 &   -0.776 \\
   0.032 &    0.636 &   -0.776 \\
   0.776 &    0.776 &    0.080 \\
\end{array} \right)
$ &
$
\left(
\begin{array}{rrr}
   0.604 &   0.0 &   -0.732 \\
  0.0 &    0.604 &   -0.732 \\
   0.732 &    0.732 &    0.004 \\
\end{array} \right)
$ 
\\
$\phantom{A}$&&&\\
$(0,\frac{1}{2},-\frac{1}{2})$ & $(0,\frac{1}{\sqrt{2}},-\frac{1}{\sqrt{2}})$ &
$
\left(
\begin{array}{rrr}
   0.068 &    0.776 &   -0.776 \\
  -0.776 &    0.664 &   -0.032 \\
   0.776 &   -0.032 &    0.660 \\
\end{array} \right)
$ &
$
\left(
\begin{array}{rrr}
  0.016 &    0.736 &   -0.736 \\
 -0.736 &    0.672 &   0.0 \\
  0.736 &   0.0 &   0.660 \\
\end{array} \right)
$ 
\\
$\phantom{A}$&&&\\
$(-\frac{1}{2},0,-\frac{1}{2})$ & $(-\frac{1}{\sqrt{2}},0,-\frac{1}{\sqrt{2}})$ &
$
\left(
\begin{array}{rrr}
   0.664 &  -0.776 &  0.032 \\
   0.776 &    0.068 &  0.776 \\
   0.032 &   -0.776 &  0.660 \\
\end{array} \right)
$ &
$
\left(
\begin{array}{rrr}
  0.672 &   -0.736 &    0.0 \\
  0.736 &    0.016 &    0.736 \\
  0.0 &   -0.736 &    0.660 \\
\end{array} \right)
$
\\
$\phantom{A}$&&&\\
$(\frac{1}{2},-\frac{1}{2},0)$ & $(\frac{1}{\sqrt{2}},-\frac{1}{\sqrt{2}},0)$ &
$
\left(
\begin{array}{rrr}
   0.636 &   -0.032 &  -0.776 \\
  -0.032 &    0.636 &    0.776 \\
   0.776 &   -0.776 &    0.080 \\
\end{array} \right)
$ &
$
\left(
\begin{array}{rrr}
   0.604 &    0.0 &   -0.732 \\
   0.0 &    0.604 &    0.732 \\
   0.732 &   -0.732 &    0.0 \\
\end{array} \right)
$
\\
$\phantom{A}$&&&\\
$(\frac{1}{2},0,-\frac{1}{2})$ & $(\frac{1}{\sqrt{2}},0,-\frac{1}{\sqrt{2}})$ &
$
\left(
\begin{array}{rrr}
  0.664 &   -0.776 &  -0.032 \\
  0.776 &    0.068 &   -0.776 \\
 -0.032 &    0.776 &    0.660 \\
\end{array} \right)
$ &
$
\left(
\begin{array}{rrr}
   0.672 &   -0.736 &   0.0 \\
   0.736 &    0.016 &   -0.736 \\
  0.0 &    0.736 &   0.660 \\
\end{array} \right)
$
\\
$\phantom{A}$&&&\\
$(0,-\frac{1}{2},-\frac{1}{2})$ & $(0,-\frac{1}{\sqrt{2}},-\frac{1}{\sqrt{2}})$ &
$
\left(
\begin{array}{rrr}
  0.068 &  0.776 &  0.776 \\
 -0.776 &  0.664 &  0.032 \\
 -0.776 &  0.032 &  0.660 \\
\end{array} \right)
$ &
$
\left(
\begin{array}{rrr}
  0.016 &    0.736 &  0.736 \\
 -0.736 &    0.672 &    0.0 \\
 -0.736 &    0.0 &    0.660 \\
\end{array} \right)
$
\\ 
$\phantom{A}$&&&\\
\hline
\hline
\end{longtable}
